\documentclass{article}

\usepackage{arxiv}

\usepackage[latin1,utf8]{inputenc}
\usepackage{hyperref}       % hyperlinks
\usepackage{amsfonts}
\usepackage{amsmath}
\usepackage{enumerate}
\usepackage{multirow}
\usepackage{graphicx}% Include figure files
\usepackage{amssymb}
\usepackage{float}% If comment this, figure moves to last Pages

\author{Roberto Dale\thanks{Corresponding author.} \\
Departamento de Estad\'istica, Matem\'atica e Inform\'atica, \\
Universidad Miguel Hern\'andez, \\
03202 Elche, Alicante, Spain \\
Centro de Investigaci\'{o}n Operativa, \\
Universidad Miguel Hern\'{a}ndez, \\
Elche, 03202 Alicante, Spain \\
\texttt{rdale@umh.es}
\And
Jes\'us M. Gand\'ia \\
Departamento de Estad\'istica, Matem\'atica e Inform\'atica, \\
Universidad Miguel Hern\'andez, \\
03202 Elche, Alicante, Spain \\
Departamento de Matemáticas, Física y Ciencias Tecnológicas, \\
Universidad Cardenal Herrera - CEU, \\
03204 Elche, Alicante, Spain \\
\texttt{jesus.martinezg@umh.es}
\And
Juan Antonio Morales-Lladosa \\
Departament d'Astronomia i Astrof\'isica,  \\
Universitat de Val\`encia, \\
46100 Burjassot, Val\`encia, Spain \\
Observatori Astron\`omic, \\
Universitat de Val\`encia, \\
46980 Paterna, Val\`encia, Spain \\
\texttt{antonio.morales@uv.es}
\And
Ramon Lapiedra \\
Departament d'Astronomia i Astrof\'isica,  \\
Universitat de Val\`encia, \\
46100 Burjassot, Val\`encia, Spain \\
Observatori Astron\`omic, \\
Universitat de Val\`encia, \\
46980 Paterna, Val\`encia, Spain \\
\texttt{ramon.lapiedra@uv.es}
}

\begin{document}

\title{\boldmath Violation of Bell inequalities from Cosmic Microwave Background data}

\maketitle

\begin{abstract}
In a recent paper [R. Dale. R. Lapiedra, and  J. A. Morales-Lladosa,  Phys. Rev. D {\bf 107}, 023506 (2023)]  a cosmic-like  Clauser-Horne-Shimony-Holt (CHSH) inequality was proved for the temperature fluctuations $\delta T$ of the perturbed Cosmic Microwave Background (CMB), assuming local realism.  This inequality can be tested from observational data. In fact, no violation for it has been found from the CMB data provided by the COBE satellite in different sky directions. This is a result which, being negative, is not conclusive in relation to the possible violation of the CHSH inequality in a cosmological context. So, the preceding analysis needs to be extended by using more precise current
CMB data, like those provided by WMAP and Planck\footnote{Based on observations obtained with Planck (http://www.esa.int/Planck), an ESA science mission with instruments and contributions directly funded by ESA Member States, NASA, and Canada. \\
\it{We dedicate this work to the memory of our colleague and friend, Professor Diego Sáez Milán,
whose insight and passion for science continue to inspire us.}} satellites. This is a task of considerable importance but, technically, much more involved in data treatments. In the present, assuming again local realism behind measurements and observational data,  this extension is accomplished.  The result is that a sound violation of the CHSH inequality is found, which would mean the failure of the assumed local realism in accordance with the quantum origin of the primordial temperature fluctuations in the framework of standard inflation and the orthodox interpretation of the Quantum Mechanics.
\end{abstract}

% keywords can be removed

\section{Introduction}
\label{sec:1}

The implementation of  Bell inequalities in a cosmic context and to test their possible violation from observational data on the sky is a current topic in Cosmology and Astrophysics. Recent studies \cite{Cam06,Mal16,MaVe17,Kan17a,Kan17b,Chou17,Che19,MaVe21,EsPo22,DaLaMo23} are concerned with this and related topics. Firstly, the research pursues the definition of Bell-like operators for quantum fields in curved space-times, using the field decomposition in Fourier modes on phase space (cf. Refs.  \cite{Cam06,Mal16,MaVe17,Kan17a,Kan17b,Chou17,Che19}). These works opened the door to analyse Bell inequalities in real space \cite{MaVe21,EsPo22,DaLaMo23} and to establish the possibility of violation from experiments which exclusively use correlations between the measured Cosmic Microwave Background (CMB) temperature anisotropies \cite{DaLaMo23}.

In addition to Bell inequalities, other indicators to display the quantum behaviour of a system or quantumness (e.g. squeezing, entanglement entropy,  mutual information, quantum discord, non-separability, etc.) are introduced in dealing with modern branches of quantum physics (e.g. optics, computation, cryptography, teleportation of quantum states, etc.). See Refs.\cite{Aul09,Hor09,Mod12} for reviews on these notions and Ref. \cite{MaMiVe23} for a comparison of different criteria for Gaussian states based on these quantumness indicators and applications. These concepts and criteria may be borrowed and prudently employed in Relativistic Cosmology scenarios to the extent that they are unambiguity defined in an expanding universe \cite{Agullo22}.  

On the other hand CMB intensity measurements, including their  fluctuations and  correlations  in different sky directions,  encode valuable information about the evolution of the observed Universe. Nowadays, there exists a board consensus among the scientific community on the certainly of the above statement (see, for instance, Ref. \cite{Pea98}). Nevertheless,  a convincing explanation (and experimental setup for the definitive  establishment)  of the statistical nature (classical or quantum) of the correlations between these fluctuations in different sky directions has not been attained yet. Really, in practice, the current observed CMB anisotropies are assumed stochastically distributed on the sky. Nevertheless, the underlying nature of the primordial fluctuations is quantum-mechanically interpreted inside the scope of the current inflationary paradigm in Cosmology \cite{Pea98,Muk81,Haw82,Gut82,Sta82,Bar83}. 

Hence, it is usually assumed that a primordial quantum distribution of cosmic fluctuations dominated a very early era of the Universe but that CMB observational data respond to a classical statistic treatment obeying a stochastic distribution \cite{Les97}. Diverse proposals exist to explain how such a quantum to classical transition takes place \cite{Les97,Nam09,Nam11,Ast20}, which are not free of criticisms \cite{Ber21}. Moreover, it is important to assess the degree of quantumness, if any, that a scalar inflationary field and the observed CMB anisotropies could encode \cite{MaVe16,Martin19}. The transition from the firstborn quantum era to a classical or semi-classical one is an important topic which continues being addressed today, both from the theoretical and observational points of views \cite{MaVe21,MaMiVe22}. Thus, the proposed theoretical models go forward to an inasmuch more interesting, but very `entangled' issue in Cosmology which claims for an experimental resolution.

The present work does not include new theoretical insight or discussion about the transition from a  quantum phase to a radiation dominant era, nor about how the primordial fluctuations were imprinted as CMB fluctuations, and its posterior evolution in an expanding universe. On the contrary, to face the issue we concentrate on the possibility to establish the true nature, if any,  of the CMB anisotropies from their own observation. We  propose a new bet to solve this `entangled' cosmic situation: to cut the `Gordian knot' with the sword of experiment. 

The present paper focuses on a variant of Bell’s inequality \cite{Bel64}, known as the CHSH inequality (named after its authors, \cite{Cla69}), whose proof has been revisited in \cite{Hes21}. This inequality has been suitably mimicked and adapted to deal with CMB anisotropies data; then it was named `cosmic CHSH inequality'. In Ref.~\cite{DaLaMo23} no violation of this inequality was reported from the analysis of the COBE mission data \cite{Cob13}. But a concluding result about the inequality fulfilment deserves fuller treatment and demands a great amount of data like those provided by the WMAP and Planck measurement results \cite{WMAP24,Planck24}.  
Now, in both cases, we find a sound violation of the Bell inequalities, which could be interpreted as a failure of local realism.
The realism assumption posits that measurements can reveal an objective physical property intrinsic to the CMB, which exists prior to and independently of any observation. The locality assumption asserts that measurements conducted in two spacelike-separated regions at the decoupling time cannot exert any physical influence on each other at any subsequent cosmic instant.

Laboratory experiments concerning the violation of Bell inequalities (see \cite{Aul09,Nobel22} and references therein) show favour to a quantum mechanical interpretation. Nevertheless it seems to us that the controversy between both quantum interpretations continues open, particularly in Cosmology \cite{Val10,Pin12}. Thus, the present paper could serve as a cosmic referee, potentially motivating the exploration of loopholes \cite{Lar-14,Asp15,Giu15,Sha15,Abe15,Lap04} and alternative classical interpretations \cite{Ber95,Ber23}, or ruling them out \cite{Gre20,Bra22}, thereby providing new insights into the longstanding debate.
This revision improves readability, precision, and fluency while maintaining a formal tone appropriate for scientific writing.

The paper is organised as follows. First, Section~\ref{sec:2} examines the fundamental hypothesis of local realism, which underpins the proof of Bell's theorem, and reconciles it with measurements of the CMB. This section also revisits the “cosmic CHSH inequality”. Section~\ref{sec:3} addresses the calculation of probabilities, expected values, and statistical methods applied to these cosmic inequalities. The computational procedures are summarised in Section~\ref{sec:4}. Subsequently, Sections~\ref{sec:5} and~\ref{sec:6} analyse observational data from the WMAP and Planck missions, demonstrating violations of the cosmic CHSH inequality in both cases. Section~\ref{sec:7} provides a concise summary of the results obtained using degraded maps, including the confirmation of these findings with the original maps. Finally, Section~\ref{sec:8} discusses the assumptions considered and the physical conclusions derived. Supplementary computations and technical results are presented in Appendices \ref{App:1} and \ref{App:2}.

\section{Cosmic CHSH inequalities}
\label{sec:2}
 
In general, the statements of the known Bell-CHSH like inequalities share in common two  essential hypothesis: physical realism and measurement locality. In the cosmic case that we are interested in, an inequality of this kind has been recently stated in Ref.~\cite{DaLaMo23}. Indeed, its proof is based on the local realism assumption, conveniently adapted to that cosmic case.  This inequality is named the cosmic CHSH inequality because it can be tested from the CMB temperature fluctuations measurements, $\delta T (\vec{x})$, in different radial directions $\vec{x}$ isotropically distributed on the sky. Concretely, the CMB measurements used in the next sections are the public results presented by the WMAP and Planck missions. 

\subsection{Local realism assumption}
\label{sec:21}

Let $A$ and $B$ denote events in a spacelike configuration at the decoupling time, and let $\boldsymbol{\hat{x}}$ and $\boldsymbol{\hat{y}}$ be unit vectors associated with past-oriented light-like directions pointing towards $A$ and $B$, respectively. Following Ref.~\cite{DaLaMo23}, we consider the dichotomised values of $\delta T (\boldsymbol{\hat{x}})$, defined by the function $F$, such that $F(\boldsymbol{\hat{x}}) = +1$ (or $-1$) if $\delta T (\boldsymbol{\hat{x}})$ is positive (or negative). A similar definition applies for the sky direction $\boldsymbol{\hat{y}}$. To derive a cosmic CHSH inequality, the hypothesis of local realism is assumed.  

By definition, realism implies the existence of hidden variables, $\lambda$, predating the measurements, such that the dichotomised CMB anisotropy values ($\pm 1$) are well-defined functions of $\lambda$ and the corresponding measurement directions. Explicitly,  
\[
\delta T_A \equiv \delta T_A(\boldsymbol{\hat{x}}, \lambda) = \pm 1, \quad 
\delta T_B \equiv \delta T_B(\boldsymbol{\hat{y}}, \lambda) = \pm 1.
\]  
As suggested by the notation, $\delta T_A$ and $\delta T_B$ are independent of $\boldsymbol{\hat{y}}$ and $\boldsymbol{\hat{x}}$, respectively. When this independence is assumed, as in the present case, the realism is said to be local.  

After each pair of measurements is completed, a new run is initiated, without requiring the subsequent pair of measurements to be conducted in the same initial state as the previous one. In a laboratory-based CHSH-like experiment, the system is actively prepared in the desired initial state before each new run. In the cosmic case, however, the invariance of the initial conditions prior to any measurement is not a crucial factor. This is because successive CMB measurement pairs are taken along randomly selected sky directions that maintain a constant angular separation. The only requirement is the assignment of a value, either $+1$ or $-1$, to each measurement.  

For each run, $\boldsymbol{\hat{x}}$ randomly assumes one of two possible values, $\boldsymbol{\hat{a}}$ or $\boldsymbol{\hat{a}'}$. Similarly and independently, 
$\boldsymbol{\hat{y}}$ takes one of two values, $\boldsymbol{\hat{b}}$ or $\boldsymbol{\hat{b}'}$. We then consider the expected values $\left\langle \delta T_A \delta T_B \right\rangle_{(\boldsymbol{\hat{x}\hat{y}})}$ for the four possible cases: $(\boldsymbol{\hat{a}\hat{b}})$, $(\boldsymbol{\hat{a}\hat{b}'})$, $(\boldsymbol{\hat{a}'\hat{b}})$, and $(\boldsymbol{\hat{a}'\hat{b}'})$. The locality assumption reflects the spacelike separation of the two measurement events during each run.  

We define the quantity  
\[
C \equiv \delta T_A (\delta T_B + \delta T_{B'}) + \delta T_{A'} (\delta T_B - \delta T_{B'}),
\]  
and the expected value $\left\langle C \right\rangle$ can be evaluated from observational data. This framework allows for the replication, in the cosmic context, of the statement and proof of the original CHSH inequalities as presented in Ref.~\cite{DaLaMo23}.

A few points should be considered regarding the terminology employed in both the standard CHSH setup and the cosmic variant explored in this work. In both contexts, the term `observer' denotes any agent (human or instrumental) capable of performing measurements and extracting the corresponding information.

In a relativistic framework, a `local observer' at an event \( A \) is represented by a future-directed timelike unit vector \( u_A \) at \( A \). Hence, any conventional laboratory test of Bell inequalities involves two space-like separated events \( A \) and \( B \), with a pair of local observers \( u_A \) and \( u_B \) stationed at each event. Each observer carries out two measurements (e.g., spin or polarisation projections) along two different spacelike directions, defined respectively by the unit vectors \( \boldsymbol{\hat{a}}, \boldsymbol{\hat{a}'} \) at \( A \) and \( \boldsymbol{\hat{b}}, \boldsymbol{\hat{b}'} \) at \( B \).

 In contrast, our proposed cosmic experiment involves a single effective observer — the satellite — performing CMB anisotropy measurements from the present epoch. However, the observed data correspond to the conditions at the time of decoupling, and thus are associated with hypothetical local observers at that earlier epoch. These observers are grouped in two pairs, \( (u_A, u_B) \) and \( (u_{A'}, u_{B'}) \), each linked to space-like separated events.

Crucially, our setup requires that all four event pairs — \( AB \), \( AB' \), \( A'B \), and \( A'B' \) — be space-like separated. During each observational run, four angular `windows' on the sky remain simultaneously open from the satellite’s perspective: \( \boldsymbol{\hat{a}}\boldsymbol{\hat{b}}, \boldsymbol{\hat{a}}\boldsymbol{\hat{b}'}, \boldsymbol{\hat{a}'}\boldsymbol{\hat{b}}, \boldsymbol{\hat{a}'}\boldsymbol{\hat{b}'} \). Each window defines a pair of events, spatially separated at the surface of last scattering, aligned along the satellite’s line of sight via the corresponding unit vectors.

Despite the conceptual differences between the usual and the cosmic CHSH setups, their experimental implementations share a key structural element: in both cases, the CHSH inequality is constructed from the outcomes of four distinct measurements, namely ${\cal O}_a$, ${\cal O}_{a'}$, ${\cal O}_b$, and ${\cal O}_{b'}$, regardless of the specific nature of the physical observable ${\cal O}$ (which may be spin, polarization, or, in our case, temperature anisotropy). In the thought-experimental version of our cosmic scenario, each observer performs a single measurement—specifically, the sign of their local temperature fluctuation: $\delta T_A$, $\delta T_{A'}$, $\delta T_B$, and $\delta T_{B'}$. At the observational level, these measurements correspond to present-day data collected by the satellite observer and derived from redshifted CMB intensity. The cosmic CHSH inequality is thus constructed from these local outcomes, under the assumption of realism—i.e., the existence of hidden variables—as outlined earlier in subsection 2.3. In this way, the structure of the standard CHSH proof can be faithfully adapted to the cosmological setting, and any observed violation of the cosmic CHSH inequality can be interpreted as evidence for the breakdown of local realism within this framework.

\subsection{Statment of the cosmic CHSH inequality}
\label{sec:22}

Let us summarise the notation and  previous definitions allowing to formulate the cosmic CHSH inequality.
\begin{enumerate}
\item [(i)] To begin with, let $\alpha \in [0, \pi]$ be the angle between $\boldsymbol {\hat x}$ and $\boldsymbol {\hat y}$, that is  $\cos\alpha \equiv {\boldsymbol  {\hat x}} \cdot {\boldsymbol  {\hat y}}$, and denote by $\left< F({\boldsymbol  {\hat x}})F({\boldsymbol  {\hat y}}),\alpha\right >$ the mean value of the product $F({\boldsymbol  {\hat x}})F({\boldsymbol  {\hat y}})$ all over the direction pairs that form a given constant angle $\alpha$. Regarding these measurement pairs, the only to be considered are the ones present in the data sets of the measures actually performed. Further, we will also use the simplified notation $\left < F({\boldsymbol  {\hat x}})F({\boldsymbol  {\hat y}}),\alpha\right > \equiv \left <\alpha\right >$.
\item [(ii)] Next we consider four different values for the constant angle $\alpha$, that is $\alpha_i$, $i = 1,2,3,4$, and the corresponding four mean values $\left < \alpha_{i} \right >$. Given these four mean values, there exist (see Ref.~\cite{DaLaMo23}) four unit vectors,  $\boldsymbol {\hat a},\, \boldsymbol {\hat a'},\, \boldsymbol {\hat b},\, \boldsymbol {\hat b'}$, such that ${\boldsymbol  {\hat a}} \cdot {\boldsymbol  {\hat b}} = \cos \alpha_{1}$, ${\boldsymbol  {\hat a}} \cdot {\boldsymbol  {\hat b'}} = \cos \alpha_{2}$, ${\boldsymbol  {\hat a'}} \cdot {\boldsymbol  {\hat b}} = \cos \alpha_{3}$, and ${\boldsymbol  {\hat a'}} \cdot {\boldsymbol  {\hat b'}} = \cos \alpha_{4}$. 

\item  [(iii)] In fact, given a quad of angles $\alpha_i$ there exist, modulo a generic rotation $R$, a two parametric quad of unit vectors $\boldsymbol {\hat a},\, \boldsymbol {\hat a'},\, \boldsymbol {\hat b},\, \boldsymbol {\hat b'}$ satisfying the required angular conditions. Thus, following the usual procedure to achieve the CHSH inequality, let us introduce a convenient algebraic sum of mean values $\left\langle C \right\rangle$ satisfying $\left\langle C \right\rangle \leq 2$ under the hypothesis of local realism. 
\end{enumerate}

\subsection{Proof of the cosmic CHSH inequality}
\label{sec:23}

According to Ref.~\cite{DaLaMo23} the cosmic CHSH inequality is formulated as:
\begin{eqnarray}
\left.\begin{aligned}
\left | \left\langle C \right\rangle\right | &\equiv\left | \left\langle F({\boldsymbol  {\hat a}})F({\boldsymbol  {\hat b}}),\alpha_{1} \right\rangle
+ \left\langle F({\boldsymbol  {\hat a}})F({\boldsymbol  {\hat b'}}),\alpha_{2} \right\rangle \right.\\
&+ \left.\left\langle F({\boldsymbol  {\hat a'}})F({\boldsymbol  {\hat b}}),\alpha_{3} \right\rangle
- \left\langle F({\boldsymbol  {\hat a'}})F({\boldsymbol  {\hat b'}}),\alpha_{4} \right\rangle
\right | \le 2.
\label{CHSH_ine2}
\end{aligned}\right.
\end{eqnarray}

For a given quad of angles, by a \textit{generic rotation} we mean a rotation that (i) maps the directions onto a set of actually observed data on the CMB sky, and (ii) excludes rotations that merely permute the observed directions among themselves.

Furthermore, the action of a generic rotation $R$ remains free so that the mean values, 
$\left< F({\boldsymbol  {\hat x}})F({\boldsymbol  {\hat y}}),\alpha\right >$, in Eq.~(\ref{CHSH_ine2})
are expressed $\left< F({R\boldsymbol  {\hat x}})F({R \boldsymbol  {\hat y}}) \right >$.
With this notation, the inequality~(\ref{CHSH_ine2}) can be written in this alternative way:
\begin{eqnarray}
\left.\begin{aligned}
\left | \left\langle C \right\rangle\right | &\equiv\left | \left\langle F(R{\boldsymbol  {\hat a}}) \left [ F(R{\boldsymbol  {\hat b}}) + F(R{\boldsymbol  {\hat b'}}) \right ] \right\rangle \right.\\
&+ \left. \left\langle F(R{\boldsymbol  {\hat a'}}) \left [ F(R{\boldsymbol  {\hat b}}) - F(R{\boldsymbol  {\hat b'}}) \right ] \right\rangle
\right | \le 2,
\label{CHSH_ineR}
\end{aligned}\right.
\end{eqnarray}
where the dependence on the angles ($\alpha_i$) remains implicit. 

Then, we can have either $F(R{\boldsymbol  {\hat b}}) = - F(R{\boldsymbol  {\hat b'}})$ or $F(R{\boldsymbol  {\hat b}}) = F(R{\boldsymbol  {\hat b'}})$.
But, because of the assumed causality, the measurement results for the directions $R{\boldsymbol  {\hat b}}$ and $R{\boldsymbol  {\hat b'}}$
do not depend on whether we measure $R{\boldsymbol  {\hat a}}$ or $R{\boldsymbol  {\hat a'}}$.
This means that in the two terms of the right hand side of inequality (\ref{CHSH_ineR}), we must put the same equality, 
$F(R{\boldsymbol  {\hat b}}) = - F(R{\boldsymbol  {\hat b'}})$ or $F(R{\boldsymbol  {\hat b}}) = F(R{\boldsymbol  {\hat b'}})$.
In each of the two cases either the first term in inequality~(\ref{CHSH_ineR}) cancels or the last term one cancels.
Therefore, the maximum value of $\left | \left\langle C \right\rangle\right |$ is two. This completes the proof of inequality (\ref{CHSH_ineR}).

The expected values $\left< F({\boldsymbol  {\hat x}})F({\boldsymbol  {\hat y}}), \alpha \right > \equiv \left < \alpha \right >$ are obtained through the expressions:
\begin{eqnarray}
\left.\begin{aligned}
\left\langle {F\left( {\boldsymbol  {\hat x}} \right)F\left( {\boldsymbol  {\hat y}} \right),\alpha } \right\rangle  &= P\left( { + , + ,\alpha } \right)  + P\left( { - , - ,\alpha } \right)  
- P\left( { + , - ,\alpha } \right) - P\left( { - , + ,\alpha } \right),
\label{Expected_V}
\end{aligned}\right.
\end{eqnarray}
where $P\left( {X,Y,\alpha } \right)$ stands for the probability of getting the result of $X\left( { \pm 1} \right)$ for $F({\boldsymbol  {\hat x}})$ and
$Y\left( { \pm 1} \right)$ for $F({\boldsymbol  {\hat y}})$ (see the $F$ function definition in Subsection~\ref{sec:21}),  for instance
$P\left( { + , + ,\alpha } \right)$ represents the probability to obtain both temperatures of the two directions subtending the angle $\alpha$
greater than the mean temperature (see Section~\ref{sec:35} for details). 

\section{Computational and statistical analysis involved in the cosmic CHSH inequalities}
\label{sec:3}

This section contains a summary of the main tasks, procedures and considerations we undertake to collect the necessary ingredients of the inequalities (\ref{CHSH_ine2}), that is the $\left< F({\boldsymbol  {\hat x}})F({\boldsymbol  {\hat y}}), \alpha \right > \equiv \left < \alpha \right >$ quantities defined in (\ref{Expected_V}). Those quantities were obtained and analysed in a previous paper \cite{DaLaMo23} using the COBE satellite data sets \cite{Cob13}. However, despite the importance of this first-generation satellite CMB measurement, two new generation satellites have improved enormously the aforementioned measures. Both satellites have significantly enhanced in terms of accuracy of the measured magnitudes as well as in terms of angular resolution. These circumstances deserve our attention and advocate a new recalculation and processing of the expected values. Once these tasks are undertaken we will review their implications.
After COBE mission, in June 2001, the Wilkinson Microwave Anisotropy Probe (WMAP) NASA Explorer mission was launched \cite{WMAP24}. While COBE satellite has an angular resolution of $7^\circ$ (FWHM -- Full width at half maximum --), the WMAP one produced the first fine-resolution of 0.2 degree supplying detailed maps. Compared to COBE mission it has 45 times the sensitivity and 33 times the angular resolution \cite{WMAP24}. Over the course of 9 years, WMAP team has produced five generation data sets named DR1, DR2, DR3, DR4 and DR5 corresponding to one, three, five, seven and nine-year data release, respectively. Finally, with a temperature resolution of one part in ${10}^6$ at an angular resolution of about 10 minutes of arc, the European Space Agency (ESA) Planck satellite \cite{Planck24} has provided the best accurate data sets. There are four public data releases labelled PR1, PR2, PR3 and PR4 relating to the released years 2013, 2015, 2018 and 2021, respectively. For an interesting comparison between both satellites (WMAP vs Planck) see \cite{Lar15}. 

\subsection{On the WMAP and Planck data sources}
\label{sec:31}

This analysis uses the last WMAP Data Release called DR5 \cite{WMAP13}. This source includes a wide range of different FITS (flexible image transport system) datafiles. For our purpose the most relevant ones are those included in the “WMAP DR5 Foreground Reduced Nine Year Coadded Sky Maps Per Frequency Band” map list. These ‘cleaned’ maps have been produced by extracting a foreground model from the `unreduced' maps using the Foreground Template Model explained in \cite{Ben13}, an enhanced algorithm of the previous one described in \cite{Hin07}. All three published Frequency Bands (Q, V and W) datafiles where considered in our calculations, the results of which are published alongside this paper. In short, those files contain a pixel label (which defines its coordinates), the Stokes I (or temperature), the Stokes Q polarisation and the Stokes U polarisation all three measurements in mK (thermodynamic). It also includes a counter field (the effective number of observations) which allows us carry out the error calculations. However, we have to clarify, that both CMB monopole and dipole have been removed from the supplied Stokes I maps. As we will see more ahead, this circumstance will be considered for the expecting values construction. 

With an improved sensitivity and resolution, a second data source obtained from the public Planck's legacy archive has been used in this study (Planck Collaboration, 2021). The Planck team has produced four different CMB maps set that have been constructed using four independent component separation pipelines, those four methods are called: COMMANDER (a Bayesian parametric method), NILC (a linear combination of data minimising a particular wavelet base), SEVEM (the CMB maps are cleaned at individual frequencies using certain templates, an then mixed to produce the final one) and SMICA (through the use of spectral covariance matrices and other parameters final maps are constructed). As it happens in the WMAP case, both monopole and dipole have been subtracted from the published Stokes I maps. The foreground cleaned maps we have used are those that belongs to the Public Data Release 3 (PR3) data set, labelled as legacy products. All these four pipelines files have been used in our calculations. In any of aforementioned files, the Stokes I, Q and U values in kelvins (thermodynamic), are supplied.

The sphere pixelisation follows the HEALPix scheme in which each pixel covers the same surface area as every other pixel \cite{Gor05}. A standard way to characterise a map resolution is through a parameter called $r$ (resolution level) which defines the number of pixels distributed on the sphere that obeys the expression  $N_{pix}=12\times N_{\rm side}^2$ where $N_{\rm side}=2^r$. In this case, the original maps we managed took $r=9$ and $r=11$ from WMAP and Planck data sets, respectively. That means that while WMAP supplies $3,145,728$ pixels per map, there are $50,331,648$ to process in the Planck case. Those quantities have to be compared with those provided by COBE satellite, that were $6,144$ pixels. The increase of these numbers, which have posed a great challenge, deserves our attention and justifies this new detailed study.

Another aspect to consider is the mask field, this is a binary field that may be supplied either in a separate file (WMAP) or as another column in the same datafile (Planck), revealing when a pixel has to be contemplated or discarded. It is designed to exclude foreground-contaminated portions of the sky. Mainly in the Galactic plane where foregrounds are dominant, but also to exclude point sources outside the Galactic plane \cite{Hin07,Planck14A12,Planck20A11}. There are two types of masks, those applied to Stokes I (temperature) and another ones to be used in polarisation studies. In the case of WMAP we used the nine-year WMAP KQ85 mask file, that excludes $18\%$ of the sky \cite{Gol09,Gje10}, while for Planck it excludes, depending on the pipeline method, a minimun of $13.2\%$ for COMMANDER and a maximun of $22.4\%$ for NILC.

\subsection{The computational challenge}
\label{sec:32}

A simple accounting of the number of pairs, without repetition, that we can obtain by combining pairs of unit vectors $\boldsymbol {\hat x}$
and $\boldsymbol {\hat y}$, that characterise our angle $\alpha$ defined by $\cos\alpha \equiv {\boldsymbol  {\hat x}} \cdot {\boldsymbol  {\hat y}}$, leads to approximately $4.95\times{10}^{12}$ for the WMAP case and $1.27\times{10}^{15}$ for the Planck one. A similar situation appears if we try to obtain the two-point correlation function or, in general, $N$-point correlation functions  \cite{Gje10,Planck20A7}. Process these huge number of pairs implies the use of unavailable computational resources or, at least, a prohibitive computational time. To solve that situation, a downgrade map resolution strategy is commonly accomplished \cite{Jar07,Gje10,Planck16,Planck20A4,Alu2023}.

Even though there are alternatives \cite{Ort24MT,Ort24}, the most common mechanism is through an average over the nearest neighbours and is carried out to degrade the Stokes I (temperature) maps and the corresponding mask fields \cite{Gje10,Planck20A7}. The aforementioned process consists of degrading the resolution and setting the value of a new superpixel as the mean of the child pixels around it. This procedure is reached integrating into our code some algorithms from the Python library Healpy \cite{Zon19}. However, a recent study claims that, in the case of two-point correlations, performing map degradation through spherical harmonic decomposition is more effective \cite{Sul24}. For this reason, we will also include in this study the results collected from maps degraded by the latter mechanism. Furthermore, in due course, we will present an evaluation of the predictions obtained from maps degraded by both methods, as they relate to our subject matter.

Regarding the mask downgrade, the neighbour-pixel averaging method yields a reduced resolution binary mask map, where each new pixel takes a value between 0 and 1. These maps are reprocessed to produce the final binary one, which is achieved by setting all pixels with values less than certain threshold to zero, while all the remaining ones have their values set to unity. The preceding threshold has been fixed to 0.50 and 0.90, in the cases of WMAP \cite{Ben13,WMAP13b} and Planck \cite{Planck16A16,Planck20A7}, respectively.

\subsection{Recovering the monopole and dipole}
\label{sec:33}
To build the final temperature map to be used in our expecting values calculations, it is necessary to recover the monopole and dipole contribution \cite{Mat99, Hin09, Jar11, Fix09, Planck20A3, Planck20A1} for each pixel. While the monopole is a fixed quantity to be added, the dipole contribution is an amplitude times the cosine of the subtended angle between observational direction and the dipole one \cite{Pee68,Mel02}. The values we have used for each satellite can be found in Table~\ref{tab:Table1}.

\begin{table}[htb]
%\begin{quote}
\caption{Monopole and Dipole values used in the expected values calculations involved in Eqs.~(\ref{CHSH_ine2})--(\ref{Expected_V}). As pointed out in \cite{Planck20A1}, the uncertainty on the Planck dipole amplitude does not include the $0.02\%$ uncertainty of the temperature of the CMB monopole.}
\label{tab:Table1}
%\end{quote}
\centering
%\resizebox{\columnwidth}{!}{
   \begin{tabular}{@{}l c c c c@{}}
 	\hline\hline
         	&  & \multicolumn{3}{c}{Dipole} \\ 
         \cline{3-5}
        	& &  & \multicolumn{2}{c}{Galactic Coordinates} \\
        	 \cline{4-5} 
        	Satellite & Monopole & Amplitude ($\mu \mathrm{K}_{\mathrm{CMB}}$) & $l$ [deg] & $b$ [deg] \\
        	\hline 
       	WMAP & $2.72548 \pm 0.00057$ K & $3355 \pm 8$ & $263.99 \pm 0.14$ & $48.26 \pm 0.03$ \\ 
        	Planck & $2.72548 \pm 0.00057$ K & $3362.08 \pm 0.99$ & $264.021 \pm 0.011$ & $48.253 \pm 0.005$ \\
        	\hline 
   \end{tabular}%}

\end{table}

\subsection{Selection criteria and sampling validation}
\label{sec:34}
The calculation of the aforementioned $\left< F({R\boldsymbol  {\hat x}})F({R \boldsymbol  {\hat y}}), \alpha \right >$ quantities for a fixed value of $\alpha$ involves a certain number of angles $\alpha_{ij}$, with $6^{\circ} < \alpha_{ij} \le 180^{\circ}$ included in a narrow interval, centred on $\alpha$, of width $\delta \alpha > 0$, that is ${\alpha _{ij}} \in \left[ {\alpha  - \delta \alpha ,\alpha  + \delta \alpha } \right]$. Notice that $\alpha_{ij} > 6^{\circ}$ is a `safety' confidence margin to discard the causal connected directions at the last scattering surface. This width $\delta \alpha$ is related to the mean spacing between pixels, defined by the map resolution. For instance to build correlation functions, WMAP established  for a $N_{\rm side}=16$ ($r=4$), which a mean spacing of $\thicksim 3^{\circ}$, a bin size of $6^{\circ}$. In that case, a total of 30 bins between $0^{\circ}$ and $180^{\circ}$ were settled \cite{Gje10}. However, since our computing capacity allows us to increase the resolution of the downgraded maps, we have degraded the original maps to a resolution of $r = 6$ (49,152 pixels per map) instead of $r = 4$ (3,072 pixels per map). Two short remarks: a resolution of $r = 6$ is the resolution we can find in Planck correlation functions papers (see \cite{Planck20A7}) and taking into account that the mean spacing, in such a case, is $\thicksim 1^{\circ}$ (bin size $\thicksim 2^{\circ}$) --see Table~\ref{tab:Table4} in Appendix~\ref{App:1}-- a total of 90 bins were adopted for numerical calculations. 

Once $\delta\alpha$ was defined, we studied the population in each bin to guarantee a $3\sigma$ confidence level. And yes, indeed, selecting a number of $10^7$ random pairs for the downgrade $r=6$ resolution maps, we fulfilled the objective. Let mention in advance that, for the original resolution of WMAP maps, that is $r=9$ (mean spacing of $\thicksim 0.12^{\circ}$), a number of $10^9$ pairs is enough. We will return to this matter later.

\subsection{Expected values calculation}
\label{sec:35}

To obtain the probabilities \( P(X, Y, \alpha) \) appearing in expression (\ref{Expected_V}), we first define the probability \( p_{\pm}(\boldsymbol{\hat{x}}) \) of obtaining the dichotomised value \( \pm 1 \) of the function \( F(\boldsymbol{\hat{x}}) \), taking into account the absolute error in the quantities \( \delta T(\boldsymbol{\hat{x}}) \). This absolute error, hereafter denoted \( \varepsilon_{a}(\delta T) \), is derived from the satellite datasets and their publicly available information (see Subsection~\ref{sec:36}). This approach corresponds to the method referred to as `Mixed Probabilities' in \cite{DaLaMo23}. A simpler alternative, known as the `Unmixed Probabilities' method, is also discussed in the same reference. A summary of the results obtained using this latter approach is provided in Appendix~\ref{App:2}.

Henceforth, for a fixed direction $\boldsymbol {\hat x}$, the $\varepsilon_{a}(\delta T)$ will represent the absolute error in $\delta T$. Based on the $\delta T$ interval defined by $[\delta T - \varepsilon_{a}(\delta T), \delta T + \varepsilon_{a}(\delta T) ]$ for a fixed direction $\boldsymbol {\hat x}$, the aforementioned $p_{\pm} (\boldsymbol {\hat x})$ are defined as follows: if we call $\varepsilon_{+}$ the positive section of the interval $[\delta T - \varepsilon_{a}(\delta T), \delta T + \varepsilon_{a}(\delta T) ]$ the probability $p_{+} (\boldsymbol {\hat x})$ can be defined as

\begin{equation}
\label{P_+}
p_{+}(\boldsymbol {\hat x}) \equiv \frac {\varepsilon_{+}} {2 \varepsilon_{a}(\delta T)},
\end{equation}
while
\begin{equation}
\label{P_-}
p_{-}(\boldsymbol {\hat x}) \equiv \frac {{2 \varepsilon_{a}(\delta T)} - \varepsilon_{+}} {2 \varepsilon_{a}(\delta T)},
\end{equation}
where, for simplicity, the $\boldsymbol {\hat x}$ dependence in $\delta T$ has been omitted.

Now, based on these expressions, we can define, for a single vector pair $(\boldsymbol {\hat x},\boldsymbol {\hat y})$ the probabilities $p_{XY}(\boldsymbol {\hat x},\boldsymbol {\hat y}) = p_{X}(\boldsymbol {\hat x}) \cdot p_{Y} (\boldsymbol {\hat y})$, which represent the probability of getting a XY value. In other words, the probability of obtaining each of the four possible values ($++$, $+-$, $-+$ and $--$ ).
 
And finally, the probabilities $P(X,Y,\alpha)$ that appear in the equation~(\ref{Expected_V}), supposing that we have a total number of $N_{\alpha}$ pairs of vectors  $(\boldsymbol {\hat x},\boldsymbol {\hat y})$ subtending certain angle $\alpha$, are
 
 \begin{equation}
\label{P_alpha_2}
P\left( {X,Y,\alpha } \right) = 
\sum\limits_{i=1}^{{N_\alpha }} {\frac{{{p_{XY}(\boldsymbol {\hat x},\boldsymbol {\hat y})_i}}}{{{N_\alpha }}}},
\end{equation}
 where $X= \pm1 $, $Y= \pm 1$. That completes the main ingredients for the calculations of the expected values (\ref{Expected_V}). Nevertheless in the next section, a few details concerning the error calculation will be discussed.

\subsection{Notes on error calculation}
\label{sec:36}
Broadly speaking, our calculation of errors is based on the standards used for propagation of uncertainty \cite{Ku66} and statistics \cite{Roh15}.
However, we will now relate a series of interesting details providing an idea of how these calculations have been carried out.

Regarding the calculation of uncertainty with WMAP measurements, we used one of the fields provided in the files containing the measurements: the effective number of observations $N_{\rm obs}$ for each pixel. From this value, we obtained pixel noise for each observation direction through the expression ${{{\sigma _0}} \mathord{\left/
 {\vphantom {{{\sigma _0}} {\sqrt {{N_{{\rm{obs}}}}} }}} \right.
 \kern-\nulldelimiterspace} {\sqrt {{N_{{\rm{obs}}}}} }}$. The $\sigma_0$ value for each frequency band can be found in \cite{WMAP13c}.

This method is valid for the original maps; however, for degraded maps, it is necessary to follow the procedure described in Appendix D of \cite{Ben13}. In our case, we calculated the uncertainty using a new effective number of observations, as given in expression D9 in Appendix D of \cite{Ben13}. Note that our type of degradation, from $r=9$ to $r=6$, is consistent with the example developed there. Additionally, we included the $0.2\%$ calibration error of WMAP \cite{Jar11,Ben13} in quadrature.

On the other hand, to calculate the errors in the case of the data provided by the Planck satellite team, we must take into account various factors such as: the accuracy of the measuring instruments (temperature resolution), systematic errors, calibration, etc. Taking these considerations into account, we set the relative error in the original measurements at $0.1\%$ level \cite{Planck11A2,Planck14A1,Planck16A3}.

It is worth mentioning that, in addition to the errors described above, that affect the data collected directly from the maps, those that appear in 
Table~\ref{tab:Table1} have to be implemented.

Once we have the error in each of the directions, it is immediate to obtain the absolute error in $\delta T(\boldsymbol {\hat x})$, that is the $\varepsilon_{a}(\delta T)$ quantities.

With regard to the errors for each of the four probabilities $XY$ for a certain angle $\alpha$ that we denote by $P(X,Y,\alpha)$, it can be retrieved from the standard statistics expression \cite{Roh15}

\begin{equation}
\label{Error_P_alpha}
{\varepsilon _{XY}}\left( \alpha  \right) = Z\sqrt {\frac{{P\left( {X,Y,\alpha } \right)\left[ {1 - P\left( {X,Y,\alpha } \right)} \right]}}{{{N_\alpha }}}},
\end{equation}
where $Z$ stands for $z$-score whose chosen value was $Z=1.96$ at $2\sigma$ level. From these values we can calculate the absolute error in the expected values
defined in (\ref{Expected_V}), by means of
\begin{equation}
\label{Error_VE_alpha}
{\varepsilon _a}\left[ \left\langle \alpha \right\rangle \right] = {\left( {{\varepsilon _{ +  + }}^2 + {\varepsilon _{ +  - }}^2 + {\varepsilon _{ -  + }}^2 + {\varepsilon _{ -  - }}^2} \right)^{{\textstyle{1 \over 2}}}}.
\end{equation}

The last step that remains for us, to complete the error calculation, is the one corresponding to the magnitude $\left| {\left\langle C \right\rangle } \right|$ detailed in
the inequality  (\ref{CHSH_ine2}). Proceeding in the same way as in our previous step, we have finally
\begin{eqnarray}
\left.\begin{aligned}
{\varepsilon _a}\left( {\left| {\left\langle C \right\rangle } \right|} \right)  &= \left [ {\varepsilon _a}^2\left( {\left\langle {{\alpha _1}} \right\rangle } \right) +
{\varepsilon _a}^2\left( {\left\langle {{\alpha _2}} \right\rangle } \right) + {\varepsilon _a}^2\left( {\left\langle {{\alpha _3}} \right\rangle } \right) \right.
+ \left. {\varepsilon _a}^2\left( {\left\langle {{\alpha _4}} \right\rangle } \right)
\right ]^{{\textstyle{1 \over 2}}}.
\label{Error_C}
\end{aligned}\right.
\end{eqnarray}
 
\section{Summary of computing tasks}
\label{sec:4}

In this section, we provide a concise review of the computational techniques employed in obtaining the results related to our evaluation of the CHSH inequalities (\ref{CHSH_ine2}). Before presenting the final outcomes, we will outline key details of the computational process, which was implemented in several stages.

\begin{enumerate}[(i)]
\item To downgrade maps from an initial resolution of $r=9$ for WMAP and $r=11$ for Planck to a final resolution of $r=6$ in both cases, we developed our own code utilising standard libraries from Healpy \cite{Zon19}. This code performs the degradation process using two distinct methods, as outlined in Subsection~\ref{sec:34}, thereby producing two degraded versions of each original map. Regarding the methods employed:  
\begin{enumerate}[(a)]  
\item The method we refer to as the `average of neighbours' assigns the value of each new superpixel as the mean of its neighbouring child pixels.  
\item The second method, based on the harmonic space approach described in \cite{Planck20A7} and \cite{Mui18}, involves spherical harmonic decomposition to extract components. From these components, we calculate the harmonic space representations of the Gaussian beam and pixel window functions, corresponding to the full width at half maximum (FWHM) and pixel resolution, respectively, for both the input and output maps. By combining these elements, we derive the downgraded harmonic coefficients, which are subsequently used to reconstruct the map at the desired resolution.  
\end{enumerate}  

\item A second code processes a total of seven downgraded maps, three from WMAP and four from Planck. This algorithm calculates the expected values, considering the elements we have described in the previous sections. In order to optimise the computation times, this code performs the operations based on a triangular matrix $49,152 \times 49,152$ (total number of pixels for the resolution $r=6$) containing all the possible angles $\alpha$. Note that as the resolution increases, this technique based on the aforementioned triangular matrix is no longer viable, since for example, for a resolution $r=9$, such as the original WMAP resolution, more than $70,000$ GB of RAM would be necessary. An outrageous amount of memory, beyond our reach.

As we mentioned in Section~\ref{sec:34}, taking into account that for this resolution the beam size is $\thicksim 2^{\circ}$, we obtained, for each map, a total of 90 expected values from which we discarded those corresponding to $\alpha_{ij} \le 6^{\circ}$.

\item For all the different feasible expected values analysed, a scanning code procedure, whose objective of finding four $\alpha$’s values which maximised function $\left | \left\langle C \right\rangle\right |$, was written. Each feasible set of four angles $\left\{ {{\alpha _1},{\alpha _2},{\alpha _3},{\alpha _4}} \right\}$ must satisfy the inequalities (see Ref.~\cite{DaLaMo23})

\begin{subequations}
\label{alphai_cond}
\begin{equation}
\label{alphai_cond_a}
\left |{\cos \alpha_{3} -\cos \beta \cos \alpha_{1}} \right | \le {\sin \beta \sin \alpha_{1}},
\end{equation}
and
\begin{equation}
\label{alphai_cond_b}
\left | {\cos \alpha_{4} -\cos \beta \cos \alpha_{2}} \right | \le {\sin \beta \sin \alpha_{2}}.
\end{equation}
\end{subequations}

Thus, for each candidate combination we look for a value for the free parameter $\beta$ that honours these conditions. To accomplish this, a loop routine looks for it. If a $\beta$ value is found, the four-angle set is approved to be processed otherwise it will be discarded.

\item \label{itm:iv} Finally all the acceptable four-set were computed, and the $\left | \left\langle C \right\rangle\right |$ were collected. A matrix containing the twenty-five largest values of $\left | \left\langle C \right\rangle\right |$, as well as the corresponding set of four angles and the respective errors for further analysis, was constructed and stored. 
\end{enumerate}

\begin{figure*}[ht]
\begin{center}
\includegraphics[scale=0.60]{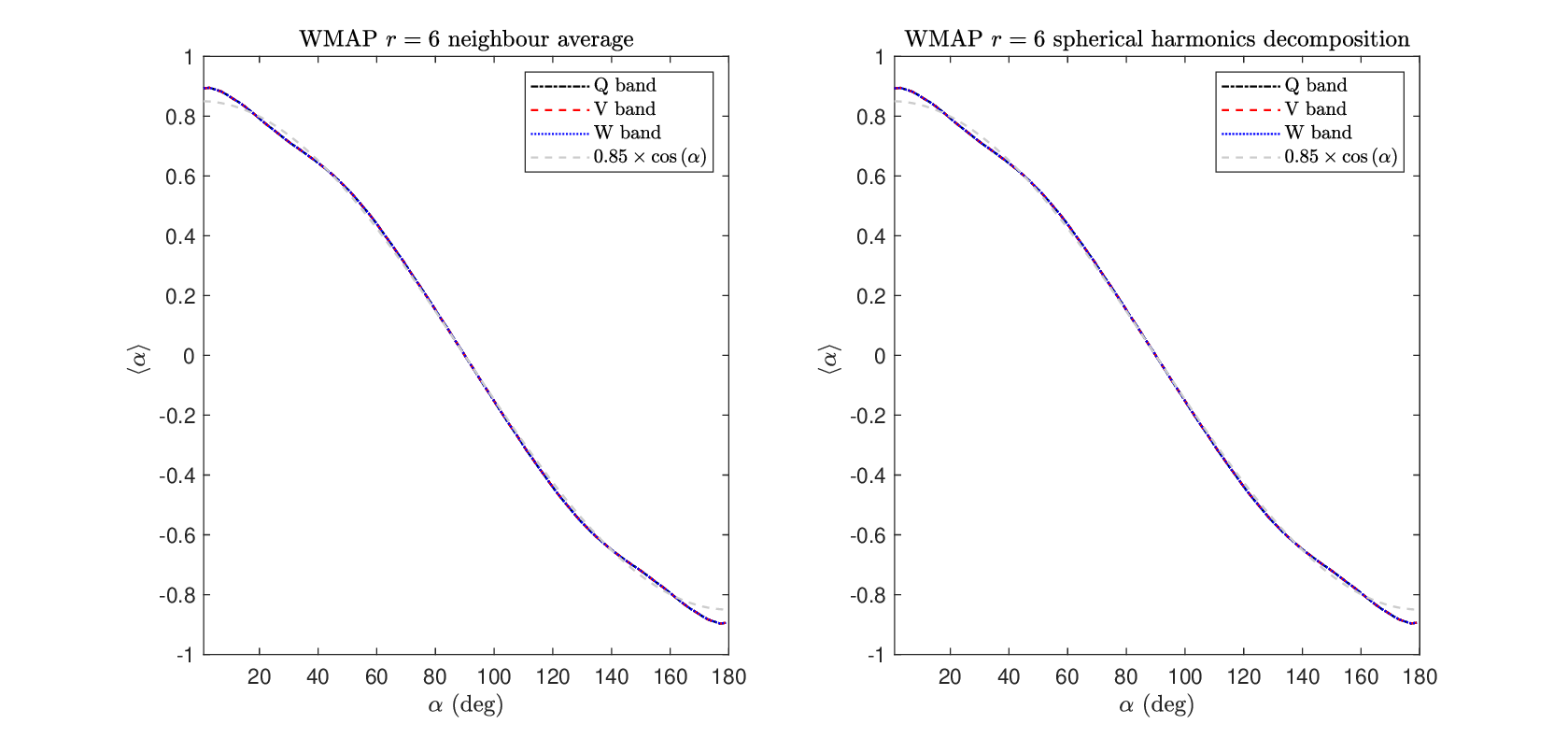}
\end{center}
\caption{This two-panel figure depicts the curves $\left\langle \alpha \right\rangle \equiv \left\langle {F\left( {\hat x} \right)F\left( {\hat y} \right),\alpha } \right\rangle$ corresponding to the expected values that have been calculated from the WMAP Q (dashdotted black line), V (dashed red line) and W (dotted blue line) band maps degraded at a resolution $r=6$, for a bean width of $2^{\circ}$ degrees. In the left panel, the degradation method by averaging neighbouring pixels has been used, while in the right panel, such degradation has been performed by spherical harmonic decomposition.
The obtained curve $\left\langle \alpha \right\rangle$ is well approximated by the function $0.85 \, \cos \left( \alpha \right)$ (dashed gray line).}
\label{fig:Figure_01}
\end{figure*}

\section{NASA-WMAP summarised outcomes}
\label{sec:5}										% OLD sec:42

Prior to presenting the final results in a series of figures and tables, it is essential to note that the selection of $10^7$ measurements within each beam, for $r=6$, has been conducted randomly one hundred times. Furthermore, the results presented below correspond to the most unfavourable case, namely, the lowest values obtained for the maximum of $\left| {\left\langle C \right\rangle } \right|$ across the one hundred random scenarios.

\subsection{The expected values $\left < \alpha \right >$}
\label{sec:51}

In Figure~\ref{fig:Figure_01}, the expected value curves (\ref{Expected_V}) derived from the analysis are presented in two panels, corresponding to the distinct downgrade methodologies employed. At first glance, two noteworthy properties can be emphasised: (i) the three curves, representing the Q, V, and W bands, are nearly identical, and (ii) consistent with results reported in analogous studies utilising data from the COBE satellite \cite{DaLaMo23}, the curves exhibit a pattern resembling a weighted cosine function, closely approximating the form $A \cos \left( \alpha \right)$. The most significant deviations from this form are observed for angles $\alpha \lesssim 18^{\circ}$ and $\alpha \gtrsim 165^{\circ}$ degrees.
However, while the approximate amplitude reported in the COBE study was $0.60$, the amplitude observed in this analysis is notably higher, around $0.85$. This discrepancy motivated us to conduct a detailed investigation, thoroughly re-examining the potential violation of the (\ref{CHSH_ine2}) inequalities.

Specifically, for the degraded maps obtained using the neighbour averaging method, the mean relative difference in expected values between the different bands is $0.06\%$, with the maximum of $0.27\%$ observed between the Q and W bands at $\alpha = 91^{\circ}$, and a minimum of $0.00005\%$ between the V and W bands. On the other hand, for the maps downgraded via spherical harmonics, the average relative difference is $0.05\%$, with a maximum of $0.16\%$ (Q and W bands) at $\alpha = 95^{\circ}$, and a minimum of $0.0003\%$ (V and W bands).

\begin{table}[ht]
\centering
\caption{This table presents the five highest values of the $\left| {\left\langle C \right\rangle } \right|$ function. Results are shown for maps degraded using both the neighbour averaging and spherical harmonic decomposition methods. For each type of map, the corresponding cases for the Q, V, and W frequency bands are included. The table lists the values of the angles that comprise the four-angle set, the $\left| {\left\langle C \right\rangle } \right|$ values, and the associated errors for each case. All angles are expressed in degrees.}
\label{tab:Table2}
%\resizebox{\columnwidth}{!}{
\begin{tabular}{ccccccccc}
\hline
\multirow{2}{*}{\begin{tabular}[c]{@{}c@{}}Downgrade\\ Method\end{tabular}} & \multicolumn{4}{c}{Four-angle set} & Q Band & V Band & W Band & Q, V and W \\ \cline{2-9} 
 &
  $\alpha_1$ &
  $\alpha_2$ &
  $\alpha_3$ &
  $\alpha_4$ &
  $\left| {\left\langle C \right\rangle } \right|$ &
  $\left| {\left\langle C \right\rangle } \right|$ &
  $\left| {\left\langle C \right\rangle } \right|$ &
  ${\varepsilon _a}\left( {\left| {\left\langle C \right\rangle } \right|} \right)$ \\ \hline
\multirow{5}{*}{\begin{tabular}[c]{@{}c@{}}Neighbour \\ Average\end{tabular}} 
								     & 133     & 135     & 137    & 45    & 2.4248 & 2.4262 & 2.4263 & 0.0009     \\
                                                                               & 133     & 135     & 139    & 47    & 2.4234 & 2.4248 & 2.4249 & 0.0009     \\
                                                                               & 133     & 137     & 139    & 49    & 2.4208 & 2.4222 & 2.4224 & 0.0009     \\
                                                                               & 133     & 135     & 141    & 49    & 2.4193 & 2.4207 & 2.4208 & 0.0009     \\
                                                                               & 131     & 135     & 141    & 47    & 2.4188 & 2.4202 & 2.4203 & 0.0009     \\ \hline
\multirow{5}{*}{\begin{tabular}[c]{@{}c@{}}Spherical\\ Harmonics\end{tabular}} 
								     & 133     & 135     & 137    & 45    & 2.4165 & 2.4183 & 2.4194 & 0.0009     \\
                                                                               & 133     & 135     & 139    & 47    & 2.4153 & 2.4171 & 2.4182 & 0.0009     \\
                                                                               & 133     & 137     & 139    & 49    & 2.4129 & 2.4146 & 2.4157 & 0.0009     \\
                                                                               & 133     & 135     & 141    & 49    & 2.4115 & 2.4132 & 2.4143 & 0.0009     \\
                                                                               & 131     & 135     & 141    & 47    & 2.4108 & 2.4126 & 2.4137 & 0.0009     \\ \hline
\end{tabular}%}
\end{table}

\begin{figure*}[ht]
\begin{center}
\includegraphics[scale=0.60]{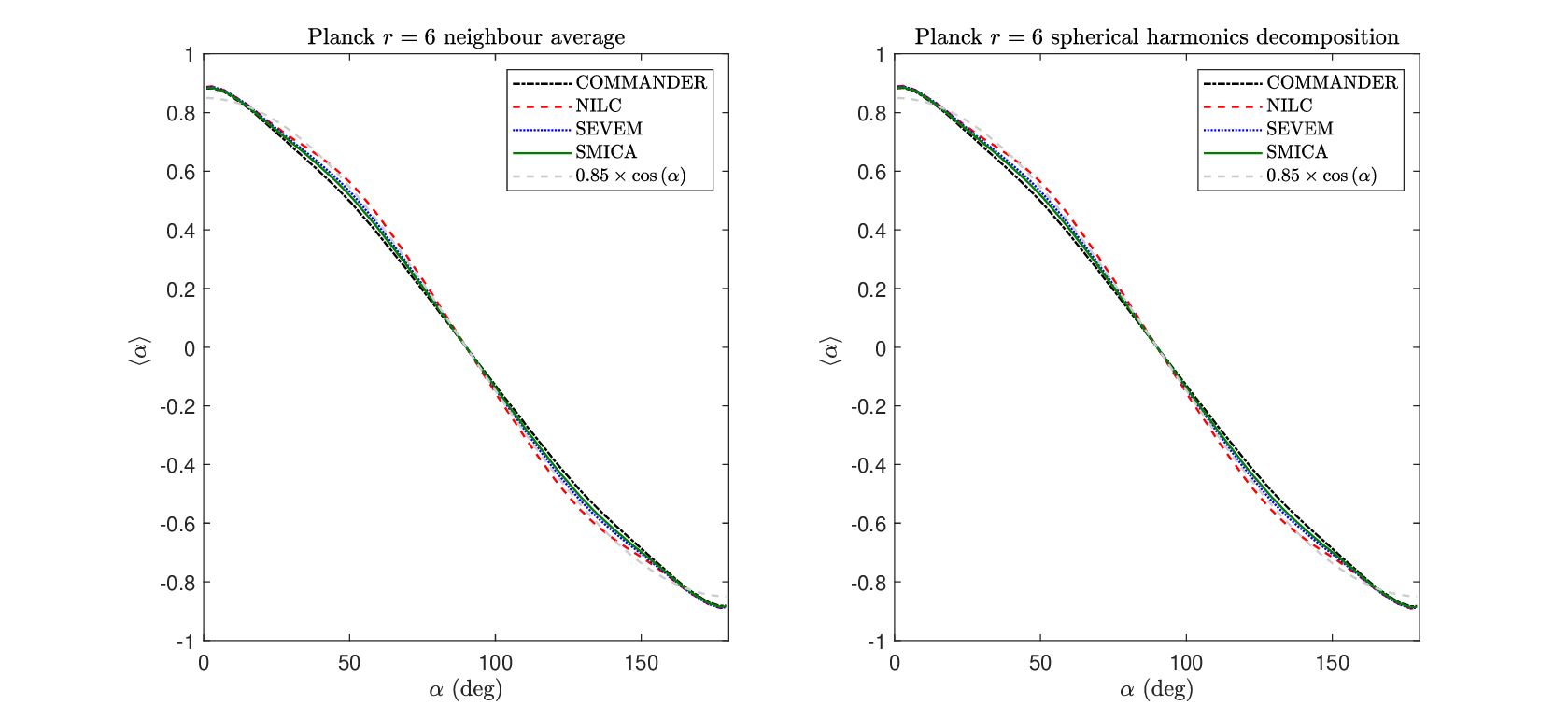}
\end{center}
\caption{As shown in Figure~\ref{fig:Figure_01}, the left and right panels present curves of the expected value $\left < \alpha \right >$ calculated from Planck maps degraded using the average of neighbours method and spherical harmonic expansion, respectively. In both cases, four curves are plotted (one for each pipeline: COMMANDER, dash-dotted black line; NILC, dashed red line; SEVEM, dotted blue line; SMICA, solid green line), along with a fifth curve representing the $0.85 \, \cos \left( \alpha \right)$ function (dashed gray line).}
\label{fig:Figure_02}
\end{figure*}

\subsection{Evaluating the quantities $\left| {\left\langle C \right\rangle } \right|$}
\label{sec:52}

Once the results of the expected values $\left\langle \alpha \right\rangle$ have been presented, for each of the Q, V and W bands, in the two types of maps, according to the degradation method, we summarise in Table~\ref{tab:Table2} the five best results, concerning the $\left| {\left\langle C \right\rangle } \right|$ values. In other words, in Table~\ref{tab:Table2} we present a summary of the five highest values obtained for the quantity $\left| {\left\langle C \right\rangle } \right|$, according to the frequency band and the degradation method of the map. This is achieved through the procedure described in Item~(\ref{itm:iv}) of Section~\ref{sec:4}.

From the results obtained, the following points merit attention:

\begin{enumerate}[(a)]
\item In the twenty-five stored cases, of which only five are presented in Table~\ref{tab:Table2}, Bell's cosmic inequalities are violated by approximately 
$20\%$.
\item As observed in Table~\ref{tab:Table2}, the five \textit{best} tetrads of angles (hereinafter referred to simply as tetrad or tetrads) are the same for all frequency bands, as well as for both map degradation mechanisms.
\item In all tetrads, the following hierarchy of expected values is observed: $\left| {{{\left\langle C \right\rangle }_Q}} \right| < \left| {{{\left\langle C \right\rangle }_V}} \right| < \left| {{{\left\langle C \right\rangle }_W}} \right|$, where the subscripts denote the Q, V, and W bands, respectively. The largest variations within a single tetrad, when comparing different frequency bands, are on the order of $\sim 0.06\%$ (between the Q and W bands). In contrast, the differences are negligible, approximately $\sim 0.006\%$, when comparing the V and W bands.
\item Similarly, when comparing results across different map degradation types, the differences (for the same band and tetrad) are approximately $\sim0.3\%$. \end{enumerate}

\section{ESA-Planck summarised outcomes}
\label{sec:6}								% Nova secció

Continuing with the presentation of results, we now summarise the findings obtained from the four types of Planck maps, which correspond to the four data processing pipelines.

\subsection{The expected values $\left < \alpha \right >$}
\label{sec:61}

In this case, Figure~\ref{fig:Figure_02} displays the expected value curves $\left < \alpha \right >$ calculated using the same two degradation methods previously applied to WMAP data, for each of the aforementioned processing pipelines. Several general characteristics can be observed in the figure: (i) unlike the curves in Figure~\ref{fig:Figure_01}, where it is difficult to visually discern which curve corresponds to each band, in this instance, the differences between the pipelines are clearly distinguishable, and (ii) as was the case with WMAP, the curves exhibit a good fit to the function $0.85 \, \cos \left( \alpha \right)$. However, while, as with WMAP, noticeable discrepancies are observed between the four curves and the referenced function within a small range at the start and end of the curves, additional subtle but discernible differences are also present in other regions, depending on the specific curve under consideration.

\subsection{Comparison of Expected Values from Different ESA-Planck Pipelines}
\label{sec:62}								% Nova Subsecció

\begin{figure}[ht]
\begin{center}
\includegraphics[scale=0.60]{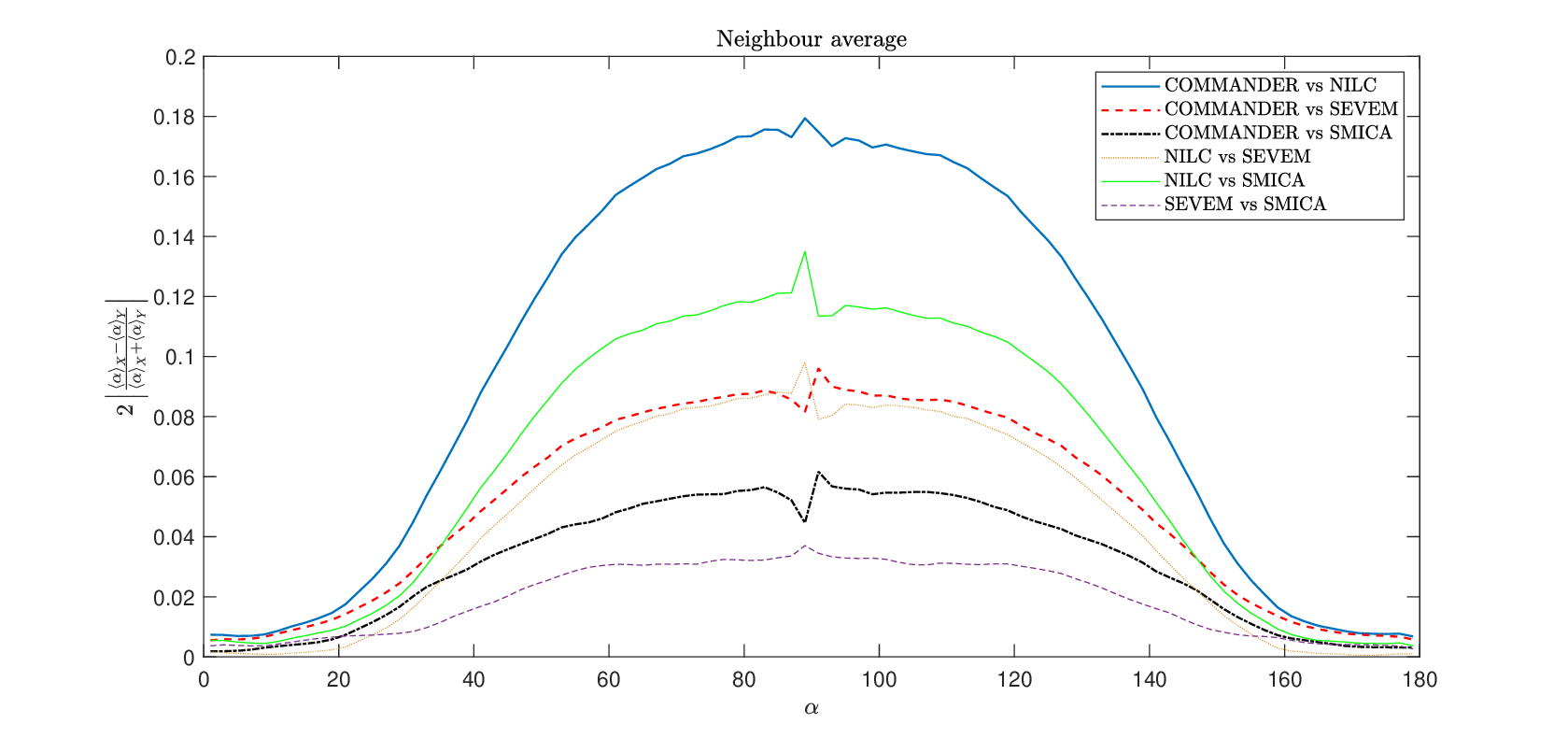}
\end{center}
\caption{In this figure we present the relative differences that we have called $\Delta_{XY}$ and that reflect the relative differences in the expected values $\left\langle \alpha  \right\rangle$ calculated from the Planck maps provided by the subsequent methodology (COMMANDER. NILC, SEVEM and SMICA).  If we combine the possible cases two by two without repetition, we obtain the six curves that we have included here. Depending on the pair compared, we have used a different color and type of plot as indicated in the legend box in the upper right corner.}
\label{fig:Figure_03}
\end{figure}

To quantitatively assess the differences observed qualitatively between the pipelines, Figure~\ref{fig:Figure_03} presents the relative differences, defined as 

\begin{equation}
\label{Delta_XY}
\Delta_{XY} \equiv 2 \left| \frac{{\left\langle \alpha \right\rangle_X - \left\langle \alpha \right\rangle_Y}}{{\left\langle \alpha \right\rangle_X + \left\langle \alpha \right\rangle_Y}} \right|,
\end{equation}
where $X$ and $Y$ represent any two distinct pipelines. 

As the curves corresponding to the two degradation mechanisms for the same pipelines are nearly identical and indistinguishable by eye, only the curves for map degradation via the average of neighbouring pixels are shown. Numerically, the mean of all these minimal differences is approximately $0.8\%$.

From Figure~\ref{fig:Figure_03}, it is evident that the pattern of differences between pairs follows a similar trend, with the maximum occurring around $\left[ 80^{\circ}, 91^{\circ} \right]$ in all cases. The largest differences are observed between the COMMANDER and NILC methods, reaching a maximum of $18\%$, while the smallest differences are between the SEVEM and SMICA methods, with a peak difference of $3.7\%$. These results suggest that there will be notable variations in the values of $\left| \left\langle C \right\rangle \right|$, which will be addressed next.

\begin{table*}[]
\caption{Following the approach used for the WMAP data in Table~\ref{tab:Table2}, this table presents the five highest values of $\left| {\left\langle C \right\rangle } \right|$ and their corresponding absolute errors, calculated from data provided by the Planck mission. These values are calculated from Planck maps degraded to resolution $r=6$. The data are organised according to the map degradation methodology applied in each of the four pipelines that generated the Planck foreground-cleaned CMB maps. For each pipeline, the tetrads are ordered in descending order of $\left| {\left\langle C \right\rangle } \right|$. All angles are expressed in degrees.}
\label{tab:Table3}
\resizebox{\textwidth}{!}{ % Ensures it uses the full width of both columns
\begin{tabular}{ccccccccccccc}
\hline
\multirow{3}{*}{Pipeline}  & \multicolumn{12}{c}{Downgrade Method}                                                                                                                                                                                                                                                                                                                                                                                                                               \\ \cline{2-13} 
                           & \multicolumn{6}{c}{Neighbour Average}                                                                                                                                                                                           & \multicolumn{6}{c}{Spherical Harmonics}                                                                                                                                                                                          \\ \cline{2-13} 
                           & $\alpha_1$ & $\alpha_2$ & $\alpha_3$ & $\alpha_4$ & \multicolumn{1}{c}{$\left| {\left\langle C \right\rangle } \right|$} & \multicolumn{1}{c}{${\varepsilon _a}\left( {\left| {\left\langle C \right\rangle } \right|} \right)$} & $\alpha_1$ & $\alpha_2$ & $\alpha_3$ & $\alpha_4$ & \multicolumn{1}{c}{$\left| {\left\langle C \right\rangle } \right|$} & \multicolumn{1}{c}{${\varepsilon _a}\left( {\left| {\left\langle C \right\rangle } \right|} \right)$} \\ \hline
\multirow{5}{*}{COMMANDER} 
& 133  & 135  & 137  & 45   & 2.1960  & 0,0009 & 133  & 135  & 137  & 45   & 2.1966  & 0.0009\\
& 133  & 135  & 139  & 47   & 2.1950  & 0.0009 & 133  & 135  & 139  & 47   & 2.1956  & 0.0009\\
& 43   & 45   & 47   & 135  & 2.1949  & 0.0009 & 43   & 45   & 47   & 135  & 2.1956  & 0.0009\\
& 41   & 45   & 47   & 133  & 2.1937  & 0.0009 & 41   & 45   & 47   & 133  & 2.1943  & 0.0009\\
& 133  & 135  & 141  & 49   & 2.1932  & 0.0009 & 133  & 135  & 141  & 49   & 2.1938  & 0.0009\\ \hline
\multirow{5}{*}{NILC}      
& 133  & 135  & 137  & 45   & 2.4376  & 0.0010 & 133  & 135  & 137  & 45   & 2.4378  & 0.0009\\
& 133  & 135  & 139  & 47   & 2.4360  & 0.0009 & 133  & 135  & 139  & 47   & 2.4362  & 0.0009 \\
& 43   & 45   & 47   & 135  & 2.4353  & 0.0010 & 43   & 45   & 47   & 135  & 2.4356  & 0.0010\\
& 41   & 45   & 47   & 133  & 2.4332  & 0.0010 & 41   & 45   & 47   & 133  & 2.4334  & 0.0010 \\
& 133  & 137  & 139  & 49   & 2.4331  & 0.0009 & 133  & 137  & 139  & 49   & 2.4333  & 0.0010\\ \hline
\multirow{5}{*}{SEVEM}     
& 133  & 135  & 137  & 45   & 2.3233  & 0.0009 & 133  & 135  & 137  & 45   & 2.3243  & 0.0009\\
& 133  & 135  & 139  & 47   & 2.3220  & 0.0009 & 133  & 135  & 139  & 47   & 2.3229  & 0.0009\\
& 43   & 45   & 47   & 135  & 2.3216  & 0.0010 & 43   & 45   & 47   & 135  & 2.3226  & 0.0010\\
& 41   & 45   & 47   & 133  & 2.3201  & 0.0010 & 41   & 45   & 47   & 133  & 2.3211  & 0.0010\\
& 133  & 137  & 139  & 49   & 2.3193  & 0.0009 & 133  & 137  & 139  & 49   & 2.3203  & 0.0009\\ \hline
\multirow{5}{*}{SMICA}     
& 133  & 135  & 137  & 45   & 2.2755  & 0.0009 & 133  & 135  & 137  & 45   & 2.2764  & 0.0009\\
& 43   & 45   & 47   & 135  & 2.2747  & 0.0010 & 43   & 45   & 47   & 135  & 2.2756  & 0.0010\\
& 133  & 135  & 139  & 47   & 2.2740  & 0.0009 & 133  & 135  & 139  & 47   & 2.2749  & 0.0009\\
& 41   & 45   & 47   & 133  & 2.2730  & 0.0010 & 41   & 45   & 47   & 133  & 2.2740  & 0.0010\\
& 133  & 137  & 139  & 49   & 2.2716  & 0.0009 & 133  & 137  & 139  & 49   & 2.2725  & 0.0009\\ \hline
\end{tabular}}
\end{table*}

\subsection{Evaluating the quantities $\left| {\left\langle C \right\rangle } \right|$}
\label{sec:63}

Following the presentation of the curves, and as previously done for the WMAP case, a summary table of the best results is provided below, this time in the case of Planck. This table (Table~\ref{tab:Table3}) includes the five highest values of $\left| \left\langle C \right\rangle \right|$, along with their associated absolute errors, for each degradation methodology and pipeline. 

As anticipated, and in contrast to the results obtained from the WMAP data --where the maximum values of the aforementioned quantities calculated from the degraded maps of the different frequency bands are virtually identical (see Table~\ref{tab:Table2})-- the calculations based on the Planck data show that the differences between the various methods can reach $\thicksim10\%$ (NILC vs COMMANDER).

In light of the results presented in Table~\ref{tab:Table3}, the following points should be noted:

\begin{enumerate}[(a)] \item For the twenty-five stored cases, of which only the five highest values are shown in this table, and for the eight map types studied (each combining two degradation methods across the four pipelines), Bell's cosmic inequalities are violated. The magnitude of this violation varies across the pipelines, ranging from a minimum of $9\%$ in the COMMANDER pipeline to a maximum of approximately $21\%$ in the NILC pipeline, with intermediate values of around $13\%$ and $15.5\%$ for the SMICA and SEVEM pipelines, respectively.
\item For all pipelines considered, both map degradation methods produce the same five best tetrads. Moreover, the ordering of these tetrads is largely preserved across the two degradation approaches.
\item In contrast to the WMAP data, where the tetrads were identical across all three frequency bands, small discrepancies in the ordering of angular values are observed in the current analysis. For instance, the tetrads $\left\{43^{\circ}, 45^{\circ}, 47^{\circ}, 135^{\circ}\right\} \) and \( \left\{133^{\circ}, 135^{\circ}, 139^{\circ}, 47^{\circ}\right\}$ appear in reverse order in the SMICA pipeline. Nevertheless, the tetrad $\left\{133^{\circ}, 135^{\circ}, 137^{\circ}, 45^{\circ}\right\}$ consistently yields the highest value across all cases.
\item In terms of the comparison between results within the same pipeline and tetrad (but different map degradation method), the differences are consistent with those observed for WMAP, at approximately $\sim0.3\%$.
\end{enumerate}

\section{Violation of the CHSH Inequalities}
\label{sec:7}	

From the calculations of $\left| \left\langle C \right\rangle \right|$ summarised in Tables~\ref{tab:Table2} (WMAP) and \ref{tab:Table3} (Planck), several noteworthy observations can be made: (i) all values violate Bell's inequalities; (ii) the most closely matched values correspond to the W band of WMAP and the NILC pipeline of Planck, with the latter exhibiting the highest value; and (iii) the angular ranges where these violations occur are remarkably similar, spanning $\left[ {45^{\circ},49^{\circ}} \right] \cup \left[ {131^{\circ},141^{\circ}} \right]$ for WMAP and $\left[ {41^{\circ},49^{\circ}} \right] \cup \left[ {133^{\circ},141^{\circ}} \right]$ for Planck.

To fulfil our commitment to revisiting the issue concerning the quality of map degradation, we now address this topic, even though the results obtained using the two different degradation methods are very similar.

While it is true that the results show minimal variation between the two mechanisms of degradation applied to the original maps, this does not entirely eliminate concerns about potential distortions in the outcomes compared to those that would be obtained using non-degraded original maps. Although we acknowledge that these degradation techniques are standard practice in studies involving correlation functions at two or more points (\cite{Jar07,Gje10,Planck16,Planck20A4,Alu2023}), we sought to address this uncertainty by conducting a portion of the calculations with the original resolution maps.

The initial code used with the downgraded maps involved a triangular matrix that encapsulated all possible angles, $\alpha$. Without delving into excessive computational detail, we note two key aspects of the modified approach: (i) the new code employed extended loops to process each matrix row individually, rather than processing the entire matrix simultaneously, and (ii) it was parallelised. While this method incurred substantial computational time, we prioritised the angle ranges where the degraded maps revealed violations. Specifically, calculations were performed in the ranges $\left[ {36^{\circ},54^{\circ}} \right] \cup \left[ {126^{\circ},146^{\circ}} \right]$.

This approach allowed us to reduce processing time significantly. Nonetheless, the calculations still required several months of computation, the results of which are now presented in a series of tables, figures, and accompanying comments. It is important to note that, while the beam width in maps with a resolution of $r=6$ is $2^{\circ}$, it is reduced to $0.25^{\circ}$ for the original WMAP maps with $r=9$ (see Table~\ref{tab:Table4} in Appendix~\ref{App:1}). Consequently, for the same range of scanning angles, the original maps feature eight times more beams than the degraded maps, along with a greater number of samples in terms of pairs per beam.

\begin{figure*}[ht]
\begin{center}
\includegraphics[scale=0.60]{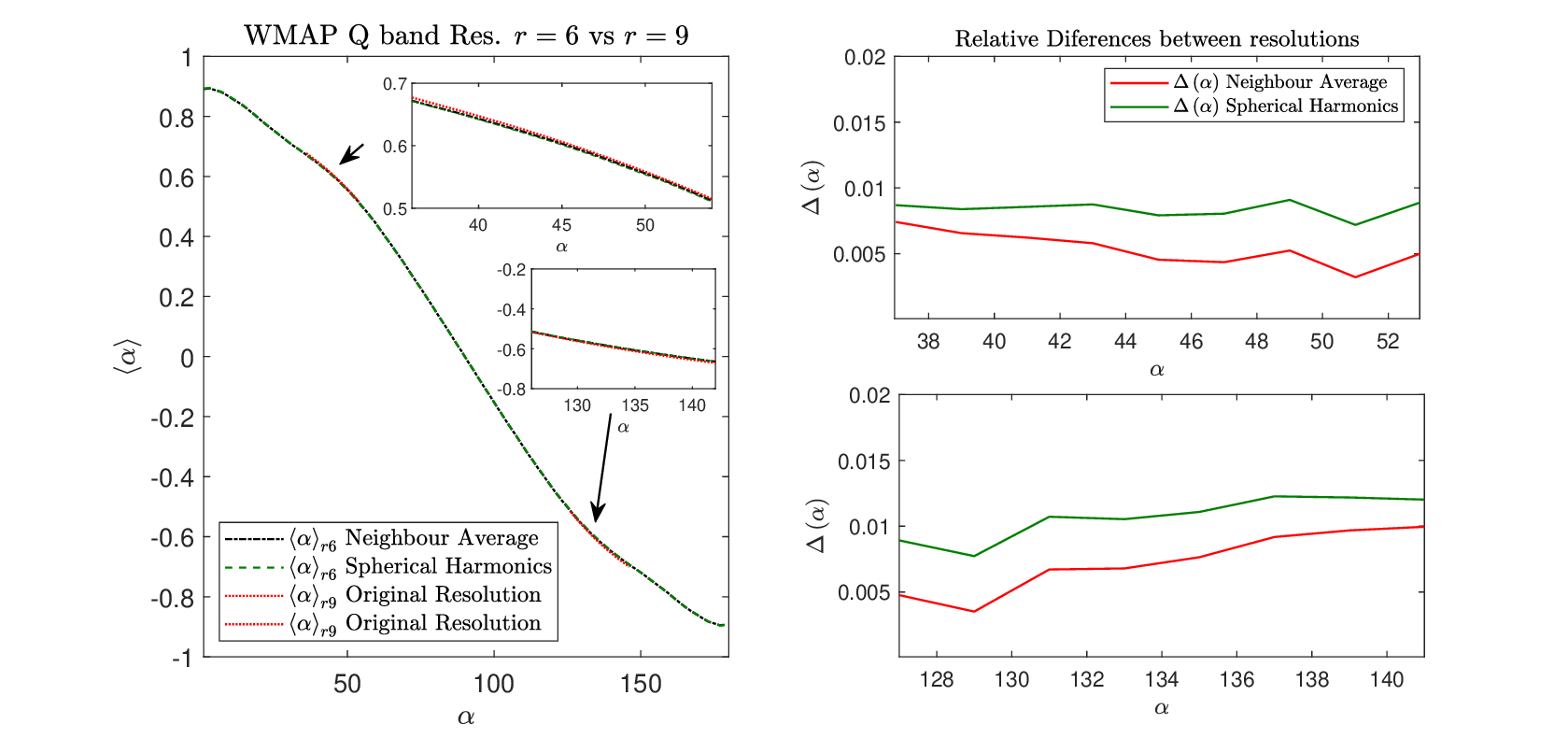}
\end{center}
\caption{This figure summarises the expected values of $\left< \alpha \right>$ from the WMAP Q band maps, displayed across three panels, comparing the downgraded resolution ($r=6$) with the original resolution ($r=9$). The left panel shows four curves: the dash-dotted black line represents the map downgraded to $r=6$ using a neighbour averaging method, the dashed green line represents the map downgraded to $r=6$ using spherical harmonics, and the two red lines correspond to the original WMAP data (at $r=9$ resolution) for the intervals $\left[ 36^{\circ}, 54^{\circ} \right]$ and $\left[ 126^{\circ}, 146^{\circ} \right]$, respectively. The two right panels illustrate the relative differences between each of the $r=6$ resolution maps and the original $r=9$ map, with the top panel corresponding to the interval $\left[ 36^{\circ}, 54^{\circ} \right]$ and the bottom panel corresponding to the interval $\left[ 126^{\circ}, 146^{\circ} \right]$.}
\label{fig:Figure_04}
\end{figure*}

Figure~\ref{fig:Figure_04} provides a summary of the comparison between the expected values $\left< \alpha \right>$ derived from the degraded maps and the original map. This comparison is presented for the case of the WMAP Q band.

To assess the differences in the expected valuess $\left< \alpha \right>$, we define the function
\begin{equation}
\label{Delta_alpha}
\Delta \left( \alpha  \right) \equiv \left| {\frac{{{{\left\langle \alpha  \right\rangle }_{r9}} - {{\left\langle \alpha  \right\rangle }_{r6}}}}{{{{\left\langle \alpha  \right\rangle }_{r9}}}}} \right|,
\end{equation}
which provides an indication of the degree of deviation between the degraded maps and the original ones. The terms ${\left\langle \alpha  \right\rangle }_{r9}$ and ${\left\langle \alpha  \right\rangle }_{r6}$ denote the expected values $\left< \alpha \right>$ for the original and downgraded maps, respectively.

Based on the collected results, as illustrated in Figure~\ref{fig:Figure_04}, a series of observations can be made regarding the expected values \( \langle \alpha \rangle \):  

\begin{enumerate}[(a)]  
\item The differences between the expected values calculated using the original and degraded maps are minimal, ranging from $0.5\%$ to $1.0\%$ for the neighbour-pixel averaging method, and from $0.8\%$ to $1.2\%$ for the spherical harmonics-based method.  

\item Throughout the studied ranges, maps degraded using the neighbour averaging method consistently produce results that are closer to those obtained from the original maps compared to those degraded through the spherical harmonics approach.  

\item The relative differences, as shown in the right-hand panels, exhibit similar patterns for both degradation methods. Moreover, the range of values is nearly identical within the two intervals (depicted in separate panels).  
\end{enumerate}  

After calculating the expected values of $\alpha$ in the original WMAP map in the Q band, we proceeded to identify the twenty-five highest values of $\left| {\left\langle C \right\rangle } \right|$. As in our previous analysis, we present below a new table (Table~\ref{tab:Table4}) showing the highest values from the aforementioned list; in this instance, we present the top ten.

\begin{table}[hb]
\centering
\caption{Following the results presented earlier for the WMAP data in Table~\ref{tab:Table2} and the Planck data in Table~\ref{tab:Table3}, this table lists the ten highest values of $\left| \langle C \rangle \right|$ together with their corresponding absolute errors. These values are derived from the non-degraded WMAP Q-band map, which has a resolution of $r = 9$. All angles are expressed in degrees.}
\label{tab:Table4}
%\resizebox{\columnwidth}{!}{
\begin{tabular}{cccccc}
\hline
\multicolumn{4}{c}{Four-angle set}                & \multicolumn{2}{c}{Expected Values}                                                                                                  \\ \hline
$\alpha_1$ & $\alpha_2$ & $\alpha_3$ & $\alpha_4$ & $\left| {\left\langle C \right\rangle } \right|$ & ${\varepsilon _a}\left( {\left| {\left\langle C \right\rangle } \right|} \right)$ \\ \hline
134.75     & 135.75      & 136.25     & 46.75     & 2,44349                                          & 0.00004                                                                           \\
134.75     & 136.00      & 136.25     & 47.00     & 2,44339                                          & 0.00004                                                                           \\
134.50     & 135.75      & 136.25     & 46.75     & 2,44335                                          & 0.00004                                                                           \\
134.75     & 135.50      & 136.25     & 46.50     & 2,44334                                          & 0.00004                                                                           \\
134.75     & 135.25      & 136.25     & 46.25     & 2,44330                                          & 0.00004                                                                           \\
134.50     & 136.00      & 136.25     & 47.00     & 2,44325                                          & 0.00004                                                                           \\
134.75     & 135.00      & 136.25     & 46.00     & 2,44324                                          & 0.00004                                                                           \\
134.25     & 135.50      & 136.25     & 45.50     & 2,44321                                          & 0.00004                                                                           \\
134.50     & 135.50      & 136.25     & 46.50     & 2,44320                                          & 0.00004                                                                           \\
134.50     & 135.25      & 136.25     & 46.25     & 2,44316                                          & 0.00004                                                                           \\ \hline
\end{tabular}%}
\end{table}

In light of the results obtained from the original maps, we can draw the following conclusions:  
(i) As observed to varying degrees across all degraded maps analysed, it is possible to identify multiple tetrads that violate the cosmic inequalities (\ref{CHSH_ine2});  
(ii) The maximum values of $\left| {\left\langle C \right\rangle } \right|$ slightly exceed those calculated for the corresponding degraded maps; and  
(iii) The angle values of each tetrad exhibit remarkable consistency across all cases.  

Moreover, random checks conducted within other narrow ranges and bands revealed similar patterns of behaviour in the results.

\section{Discussion on the made assumptions and concluding results}
\label{sec:8}

The analysis of data from the WMAP and Planck satellites regarding the CMB, using both degraded maps and samples from the original maps, has produced conclusive findings. Even in the most unfavourable scenarios considered, it is possible to identify multiple sets of four sky directions—corresponding to four pairs of causally disconnected events at the time of decoupling—that violate the cosmic CHSH inequalities (\ref{CHSH_ine2}).

The statistical and numerical analysis applied in this study demonstrates a significant violation of the cosmic CHSH inequalities, with a maximum value reaching 2.4. This result lies near the midpoint of the numerical interval $\left[2, 2\sqrt{2}\right]$, which is bounded by the classical upper limit of 2 (the quantum threshold for violation) and the maximum value of $2\sqrt{2}$ (Cirel'son limit, see \cite{Cir80}) permitted by the orthodox quantum interpretation. Such a pronounced violation suggests the breakdown of the primary assumption underpinning the derivation of the cosmic CHSH inequalities: the hypothesis of local realism. In the present cosmic context, local realism refers to the assumption that there exists a set of hidden variables, denoted by $\lambda$, predating the measurements of CMB temperature anisotropies. Under this assumption, these anisotropies, $\delta T(\boldsymbol{\hat{x}}, \lambda)$, are independent of any unit vector $\boldsymbol{\hat{y}}$ whose sky image is spacelike separated from that of $\boldsymbol{\hat{x}}$.

As in our previous studies, the analysed quantities were temperatures, although this time measured with substantially higher precision and angular resolution compared to our earlier research, which relied on data from the COBE satellite \cite{Cob13,DaLaMo23}.  

It is quite reasonable that higher maximum values of $\left| {\left\langle C \right\rangle } \right|$ were obtained when using the original maps rather than the degraded ones. Degrading a map inherently smooths out potential peaks present in the original beams, as these are sharper and more defined. Notably, the beam width for the degraded maps was $2^{\circ}$, while for the original maps, it was $0.25^{\circ}$ —making the original beams eight times narrower.

While the primary aim of this study was to evaluate CHSH inequalities through measurements of the microwave background, the experience gained in the analysis and treatment of map data enables us to draw additional conclusions. Specifically, we examined the performance of traditional map degradation methods —namely, the averaging of neighbouring pixels and the expansion in spherical harmonics— in calculating expected values. In this regard, we conclude that, at least within the angular ranges of interest for this study:  
(i) The results obtained using both degradation methods closely align with those derived from the original maps, exhibiting relative differences of only $0.5\%$ to $1\%$; and  
(ii) The analysis of maps degraded using the averaging-of-neighbours method closely approximates the results obtained from the original maps.

The present study is based on intensity data, $I$ (Stokes), provided by the WMAP and Planck satellites. After processing, these data have been utilised to examine the potential violation of the cosmic CHSH inequalities (\ref{CHSH_ine2}). As previously mentioned, $Q$ and $U$ (Stokes) polarisation data are also available. In this context, it is noteworthy that our future research efforts in this area will focus on testing Bell-like inequalities involving these polarisation data. To this end, we are currently engaged in both the processing of these data and the derivation of new inequalities.

At this stage, it is essential to examine whether certain aspects of the observational data processing or the execution of the cosmic test introduce additional assumptions that could undermine local realism. Specifically, it is necessary to assess whether the cosmic test for the CHSH inequalities effectively addresses the well-known loopholes commonly associated with such experiments, including locality, freedom of choice, and efficient detection (see Ref.~\cite{Lar-14} for a comprehensive review of this topic). First and foremost, the concept of a loophole must be adapted to the cosmic context under consideration.

We considered CMB anisotropies measured from two space-like separated regions, denoted A and B, at the time of decoupling or shortly after the inflationary epoch. These regions are not causally connected at that time. However, it is important to explore the possibility that A and B retain some form of memory of their shared causal past, represented by the intersection of their respective past light cones. This raises an intriguing question —a cosmic analogue of the conventional `freedom-of-choice' condition. Tracing back in time, one can conclude that the domain of this hypothetical interaction would be entirely contained within the inflationary epoch. Given the vast cosmic expansion during this period, it is challenging to envisage how any mutual causal connection between A and B could have persisted through their subsequent evolution. Therefore, this interpretation of the `freedom-of-choice' loophole can be reasonably circumvented in a cosmological context.

It is well known that the inequalities derived by Clauser \textit{et al.} \cite{Cla69} (CHSH inequalities) may not be violated even if local realism holds. In this context, it is necessary to assume that the intersection of the two delayed light cones corresponding to the measurements of each pair provides a common ground where the resulting correlations mimic the role of non-existent information transmission between the measurements. 

As the separation between the two measurements of each pair increases, the intersection regions of their light cones will become more distant in space and time. This has led to CHSH inequality-type experiments—spanning planetary distances in terms of the spatial separations—being conducted without revealing any violation of these inequalities (see, for example \cite{Asp15, Abe15, Lap04}).

Conversely, a recent addendum to the manuscript \cite{DaLaMo24} predicts a violation of certain cosmic CHSH inequalities. These inequalities involve pairs of measurements of primordial perturbations in cosmic energy densities at the end of inflation, taken in different directions in a space-like configuration. However, when considering the cosmological timeline, can time-like configurations give rise to space-like configurations in the future? The answer is negative: the change in configuration, driven by the development of inflation, which could account for the subsequent violation of the cosmic CHSH inequalities, is in fact impossible, since the inflationary expansion occurs at a rate faster than the expansion of cosmic matter. Nonetheless, violations of the CHSH inequalities could have occurred during the pre-inflationary period, though these effects would be undetectable today due to the extreme redshifting caused by inflation.

In summary, the eventual detection of a violation of the cosmic CHSH inequalities cannot be attributed to any potential information transmission between the measurements of a given pair, nor to signals originating from the pre-inflationary universe. Such a violation can only arise from purely quantum correlations. In our analysis, these correlations involve temperature fluctuations measured along directions on the sky corresponding to space-like separated regions at the time of last scattering. Assuming a quantum statistical description of the primordial fluctuations—consistent with the standard inflationary framework—such violations provide indirect evidence of entanglement imprinted in the early universe.

\appendix
\section{Tables and Extended Statistics.}  
\label{App:1}

In the study of CMB maps, the resolution of these maps is typically characterised by the parameter $r$ or the parameter $N_{\rm side}$. To establish a relationship between this characterisation and other properties relevant to our analysis, we present Table~\ref{tab:Table5}. The derivation of the mean pixel spacing plays a crucial role in determining the appropriate beam size for the calculation of the expected values $\left < \alpha \right >$. This calculation has been performed using algorithms implemented in the Healpy library \cite{Zon19}.  

\begin{figure}[H]
\begin{center}
\includegraphics[scale=0.60]{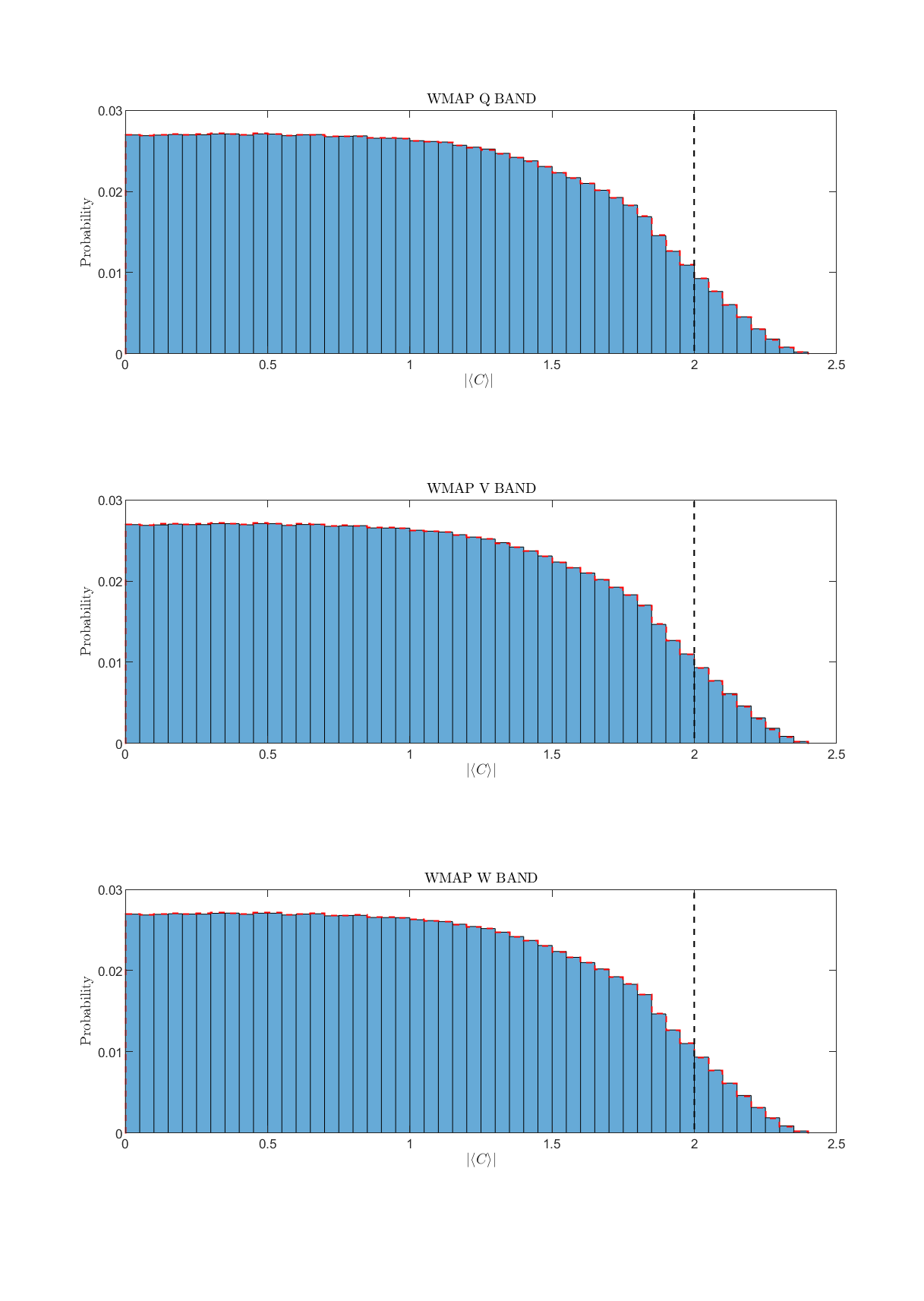}
\end{center}
\caption{This figure comprises three panels, each depicting the normalised probability distribution of all possible values of the function $\left| \langle C \rangle \right|$ for the WMAP satellite across its three frequency bands (Q, V, and W). The blue bars represent the results obtained using maps degraded via the neighbour-pixel averaging mechanism, whereas the dashed red step plot corresponds to results derived from maps degraded using spherical harmonics. The dashed black vertical line separates the sets of four directions that violate the CHSH cosmic inequalities~(\ref{CHSH_ine2}) (on the right) from those that do not (on the left). The common width used for both bars and steps in the representation is $\Delta \left| \langle C \rangle \right| = 0.05$.}

\label{fig:Figure_05}
\end{figure}

\begin{table}[h]  
\caption{Characteristic parameters of the processed maps relevant to this study. From left to right, the table presents the parameters $r$ and $N_{\rm side}$ (related via $N_{\rm side} = 2^r$), the average spacing between pixels, and the total number of pixels $N_{\rm pix} = 12 \times N_{\rm side}^2$. These parameters adhere to the HEALPix pixelation scheme \cite{Gor05}.}  
\label{tab:Table5}  
\centering
%\resizebox{\columnwidth}{!}{  
\begin{tabular}{cccc}  
\hline\hline  
$r$ & $N_{\rm side}$ & Mean Spacing (deg) & Number of Pixels \\  
\hline  
4 & 16 & 3.660 & 3,072 \\  
5 & 32 & 1.830 & 12,288 \\  
6 & 64 & 0.920 & 49,152 \\  
7 & 128 & 0.460 & 196,608 \\  
8 & 256 & 0.230 & 786,432 \\  
9 & 512 & 0.110 & 3,145,728 \\  
10 & 1,024 & 0.057 & 12,582,912 \\  
11 & 2,048 & 0.028 & 50,331,648 \\  
\hline  
\end{tabular}%}  
\end{table}

\begin{figure*}[ht]
\begin{center}
\includegraphics[scale=0.60]{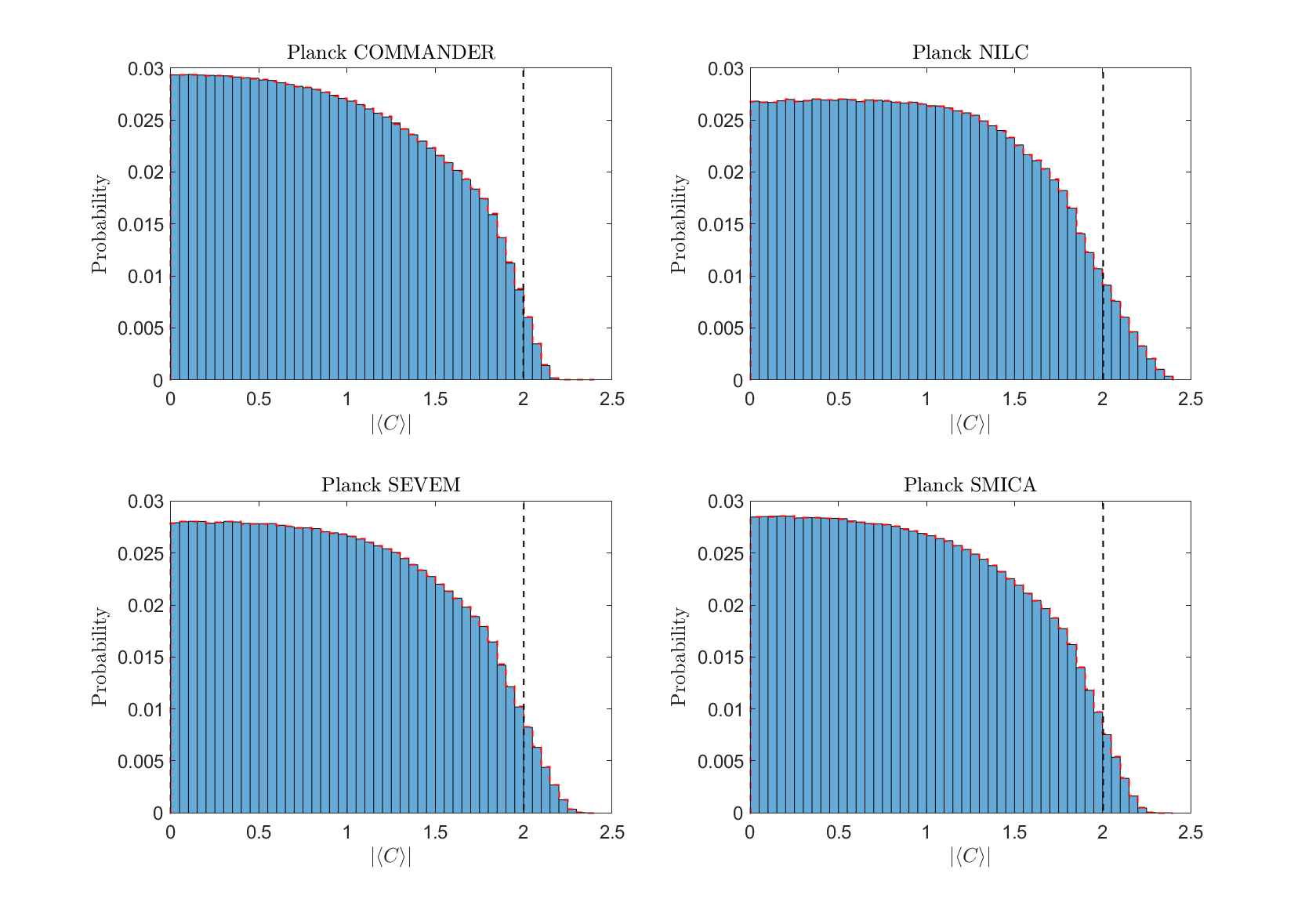}
\end{center}
\caption{As in Figure~\ref{fig:Figure_05}, the normalised probability distribution of the function $\left| \langle C \rangle \right|$ is presented, this time for the four Planck pipelines, across four panels. In each panel, the blue bars correspond to the neighbour-pixel averaging degradation method, while the dashed red step plot represents the spherical harmonics approach. The vertical dashed black line delineates the cases that violate the inequalities~(\ref{CHSH_ine2}) (to the right) from those that do not (to the left). The bin width used for both the blue bars and the red step plot is $\Delta \left| \langle C \rangle \right| = 0.05$.}

\label{fig:Figure_06}
\end{figure*}

While the primary objective of this study is to determine whether the CHSH inequalities~(\ref{CHSH_ine2}) are violated, and to this end, we have presented a series of tetrads that demonstrate such violations across all WMAP bands and Planck pipelines, additional statistical information may also be of interest to the reader. Accordingly, in this section, we provide a series of figures and tables offering a more comprehensive perspective on the results obtained for the function $\left| \langle C \rangle \right|$.

As anticipated, the distribution of values for the function $\left| \langle C \rangle \right|$ is highly consistent across the three bands of the WMAP satellite, as illustrated in detail in Figure~\ref{fig:Figure_05}. A similar pattern is observed in the percentages of cases where the inequalities~(\ref{CHSH_ine2}) are violated, as presented in the third and fourth columns of Table~\ref{tab:Table6}, corresponding to the condition $\left| \langle C \rangle \right| \geq 2$. Likewise, as expected, the probability distributions obtained from the two degradation methods—blue bars for the neighbour-pixel averaging method and red dashed steps for the spherical harmonics approach—are nearly identical and indistinguishable to the naked eye.

\begin{table}[h]
\centering
\caption{Summary of the percentage of four-direction sets for which the function $\left| \langle C \rangle \right|$ is greater than or equal to 2. This corresponds to the area to the right of the black dashed line in Figures~\ref{fig:Figure_05} and~\ref{fig:Figure_06}.}
\label{tab:Table6}
%\resizebox{\columnwidth}{!}{ % Resize to fit column width
\begin{tabular}{cccc}
\hline
\textbf{}               & \textbf{}                & \multicolumn{2}{c}{\textbf{Downgrade Method}} \\ \cline{3-4} 
\textbf{}               & \textbf{}                & \textbf{Neighbour}    & \textbf{Spherical}    \\
\textbf{Satellite}      & \textbf{Band / Pipeline} & \textbf{Average}      & \textbf{Harmonics}    \\ \hline
                        & Q                        & 3.35\%                & 3.32\%                \\
WMAP                    & V                        & 3.39\%                & 3.34\%                \\
                        & W                        & 3.39\%                & 3.35\%                \\ \hline
\multirow{4}{*}{Planck} & Commander                & 1.10\%                & 1.12\%                \\
                        & Nilc                     & 3.39\%                & 3.41\%                \\
                        & Sevem                    & 2.32\%                & 2.34\%                \\
                        & Smica                    & 1.83\%                & 1.85\%                \\ \hline
\end{tabular}%}
\end{table}

However, this broad similarity is not observed in the case of Planck across different pipelines, although it does hold when comparing the two degradation methods. A visual inspection of the four panels immediately reveals differences in the probability distributions, including: (i) variations in the maximum probabilities, (ii) noticeable differences in the areas corresponding to values of interest, specifically where $\left| \langle C \rangle \right| \geq 2$, and (iii) distinct behaviours in the probability curve within the approximate interval $\left[0,1\right]$. While this curve exhibits a similar shape for the Commander, Sevem, and Smica pipelines, it is markedly different in the case of Nilc, where it remains almost flat. In this regard, the distribution for Nilc closely resembles those observed for WMAP.

The differences in the area of interest, to the right of the vertical black dashed line, are also clearly reflected in Table~\ref{tab:Table6}, where the area fraction for Nilc matches that of the WMAP V and W bands when using the neighbour averaging degradation method. Furthermore, the table highlights a contrasting trend in the variation of the percentage with respect to the degradation method. While in the case of WMAP, the percentage slightly decreases for maps processed via spherical harmonic decomposition, the opposite trend is observed for Planck, where this percentage instead increases slightly.

\section{Unmixed Probabilities: Method and Results}  
\label{App:2}

In Sections~\ref{sec:5} and \ref{sec:6}, we have presented the results for the expected values \( \left\langle F\left( \boldsymbol{\hat{x}} \right) F\left( \boldsymbol{\hat{y}} \right), \alpha \right\rangle \), as defined in expression~\ref{Expected_V}, derived from the probabilities \( P(X, Y, \alpha) \). These probabilities were calculated using the procedure described in Subsection~\ref{sec:35}, originally introduced in \cite{DaLaMo23}, and referred to as the `Mixed Probabilities' method. This approach accounts for the absolute error intrinsic to each of the temperature measurements, such that, instead of assigning a definite value \( \pm 1 \) to \( F(\boldsymbol{\hat{x}}) \), we assign probabilities \( p_{\pm}(\boldsymbol{\hat{x}}) \) that \( F(\boldsymbol{\hat{x}}) \) takes the value \( \pm 1 \).

In this section, however, we summarise the results obtained using the so-called `Unmixed Probabilities' approach. This method is considerably simpler: for each frequency, the sign of \( \delta T \) at a given direction determines a deterministic outcome of either \( +1 \) or \( -1 \), without accounting for measurement uncertainty. As a consequence, the probabilities entering expression~\ref{Expected_V} are straightforwardly computed as:
\begin{equation}
\label{P_alpha}
P(X,Y,\alpha) = \frac{\nu(X,Y,\alpha)}{N_{\alpha}},
\end{equation}
where \( \nu(X,Y,\alpha) \) is the raw count of occurrences of the configuration \( (X,Y,\alpha) \), and \( N_{\alpha} \) is the total number of such pairings at angular separation \( \alpha \).

\begin{figure*}[h]
\begin{center}
\includegraphics[scale=0.60]{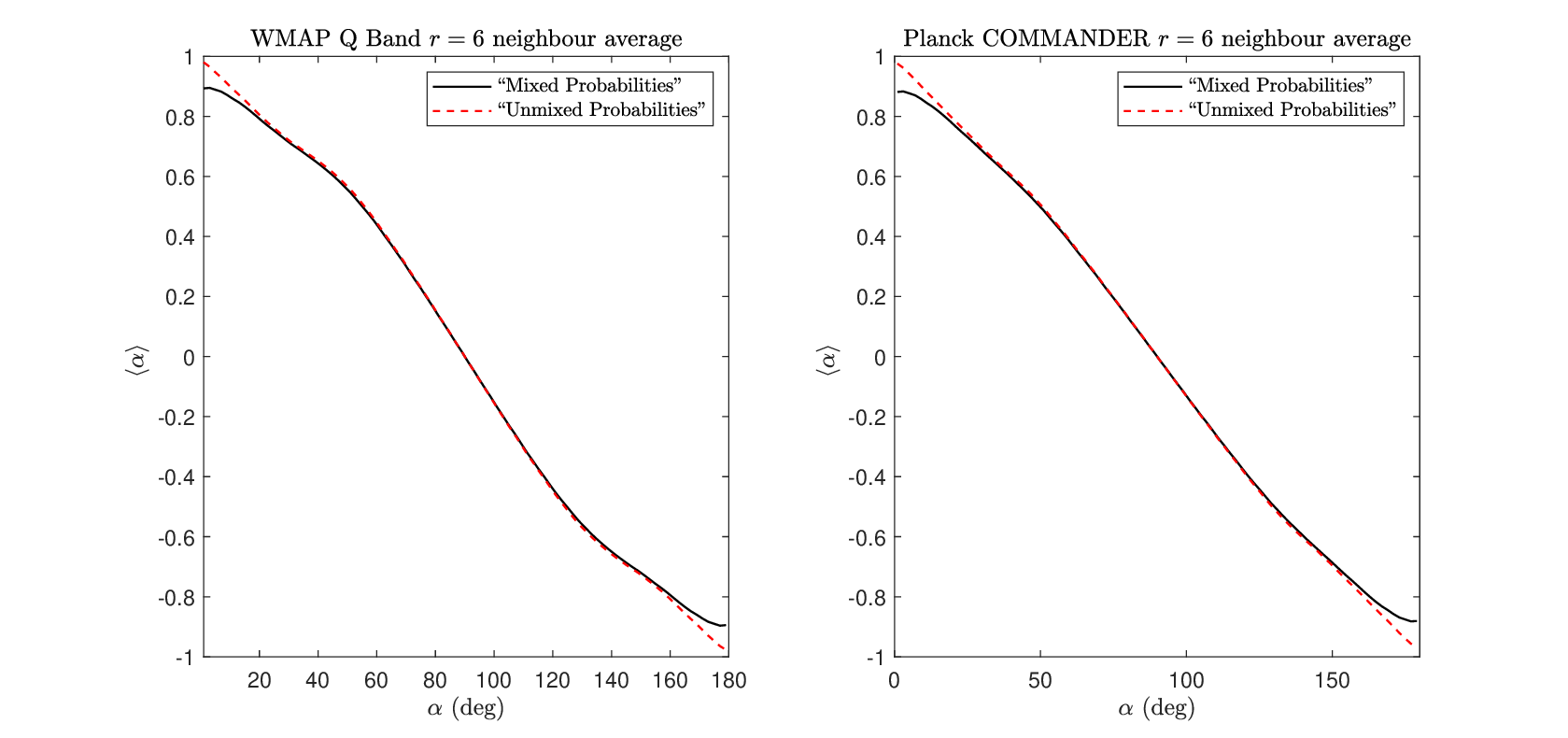}
\end{center}
\caption{Comparison between the expected values \( \left\langle F(\boldsymbol{\hat{x}}) F(\boldsymbol{\hat{x}}), \alpha \right\rangle \) calculated via the Mixed and Unmixed Probabilities methods for representative cases from each satellite mission. The solid black line represents the previously analysed case using the `Mixed Probabilities' method, while the dashed red line corresponds to the new results obtained with the `Unmixed Probabilities' approach. The left panel refers to the WMAP Q-band, while the right one refers to the Planck mission analysed using the COMMANDER pipeline.}

\label{fig:Figure_07}
\end{figure*}

In Figure~\ref{fig:Figure_07}, we show two representative cases—one from each mission—to illustrate the differences between the expected values obtained using the `Mixed' and `Unmixed Probabilities' methods. The observed behaviour is consistent across all analysed cases and may be summarised as follows:

\begin{enumerate}[(a)]
\item In the intermediate angular range \( \alpha \in [30^\circ, 150^\circ] \), there are no appreciable differences between the two methods.
\item However, as \( \alpha \to 0^\circ \) (or \( 180^\circ \)), the results begin to diverge: while the `Mixed Probabilities' curve tends towards \( +0.90 \) (or \( -0.90 \)), the Unmixed Probabilities curve asymptotically approaches the ideal values of \( +1 \) (or \( -1 \)).
\end{enumerate}

This latter result for the `Unmixed Probabilities' case is consistent with expectations. Specifically, as \( \alpha \to 0^\circ \), the unit vectors \( \boldsymbol{\hat{x}} \) and \( \boldsymbol{\hat{y}} \) become indistinguishable, and the limiting behaviour \( F(\boldsymbol{\hat{x}})F(\boldsymbol{\hat{x}}) \to F^2(\boldsymbol{\hat{x}}) = 1 \) is indeed fulfilled.

From the results obtained in the calculation of the expected values discussed above, we employed the same code to scan and identify the tetrads of angles $\left\{ \alpha_1, \alpha_2, \alpha_3, \alpha_4 \right\}$ that maximise the value of the function $\left| \left\langle C \right\rangle \right|$. In Table~\ref{tab:Table7}, we present the three highest values obtained using the Q band (WMAP mission) and the COMMANDER pipeline (Planck mission), respectively.

In light of the results presented in Table~\ref{tab:Table7}, the following remarks are in order:

\begin{enumerate}[(a)]
\item It can be affirmed that, for both the WMAP Q-band and the Planck COMMANDER pipeline, the maximum values of the function $\left| \left\langle C \right\rangle \right|$ increase by approximately $1.5\%$ compared to those previously obtained using the `Mixed Probabilities' method, as presented in Tables~\ref{tab:Table2} and \ref{tab:Table3}. This leads us to conclude that the primary scenario based on the `Mixed Probabilities' method is indeed the most conservative with respect to our claims regarding the potential violation of cosmic Bell inequalities.
  
\item The sets of angle tetrads shown in Table~\ref{tab:Table7} are identical to those previously obtained using the `Mixed Probabilities' method and reported in Tables~\ref{tab:Table2} and \ref{tab:Table3}, for both the WMAP Q-band and the Planck COMMANDER data analyses.

\item Lastly, although the ordering of these tetrads remains the same in the WMAP case, the second and third highest values are interchanged in the Planck results. However, it is important to note that the corresponding $\left| \left\langle C \right\rangle \right|$ values in these two cases are extremely close, practically indistinguishable between both probability calculation methodologies.
\end{enumerate}

\begin{table}[hb]
\centering
\caption{The three highest values of the function $\left| \left\langle C \right\rangle \right|$ obtained using the `Unmixed Probabilities' method. The left half corresponds to the WMAP Q-band, and the right half to the Planck mission using the COMMANDER pipeline.}
\label{tab:Table7}
\begin{tabular}{cccccc|cccccc}
\hline
\multicolumn{6}{c|}{WMAP Q Band} & \multicolumn{6}{c}{Planck COMMANDER Pipeline} \\ \hline
$\alpha_1$ & $\alpha_2$ & $\alpha_3$ & $\alpha_4$ & $\left| \left\langle C \right\rangle \right|$ & ${\varepsilon _a}\left( \left| \left\langle C \right\rangle \right| \right)$ & $\alpha_1$ & $\alpha_2$ & $\alpha_3$ & $\alpha_4$ & $\left| \left\langle C \right\rangle \right|$ & ${\varepsilon _a}\left( \left| \left\langle C \right\rangle \right| \right)$ \\ \hline
133 & 135 & 137 & 45 & 2.4637 & 0.0009 & 133 & 135 & 137 & 45 & 2.2321 & 0.0009 \\
133 & 135 & 139 & 47 & 2.4625 & 0.0009 & 43 & 45 & 47 & 135 & 2.2309 & 0.0009 \\
133 & 137 & 139 & 49 & 2.4600 & 0.0009 & 133 & 135 & 139 & 47 & 2.2305 & 0.0009 \\ \hline
\end{tabular}
\end{table}

\section*{Acknowledgements}
 
This work was partially funded by the Conselleria d’Educació, Universitats i Ocupació of the Generalitat Valenciana through projects CIAICO/2022/252 and ASFAE/2022/005. We gratefully acknowledge the invaluable technical assistance of our colleague Rafael Garbayo from the Computer Administration of the Statistics Department at Universidad Miguel Hernández. We also thank Professor Arjun Berera of the University of Edinburgh for his insightful comments and bibliographic information regarding our preliminary version arXiv:2302.05125.

\end{document}